\documentclass[a4paper]{article}
\usepackage[english]{babel}
\usepackage[T2A]{fontenc}
\usepackage[utf8]{inputenc}
\usepackage{amsmath}
\usepackage{amsthm}
\usepackage{amssymb}
\usepackage{textcomp}
\usepackage{graphicx}
\usepackage{array}
\usepackage{hyperref}

\newcommand{\be}{\begin{eqnarray}}
\newcommand{\ee}{\end{eqnarray}}

\newcommand{\eqlb}[2]{\begin{equation} \label{#1} #2 \end{equation}}
\newcommand{\eq}[1]{\begin{equation} #1 \end{equation}}

\newcommand{\brc}[1]{\left(#1\right)}

\newcommand{\bfi}[1]{\left\{ #1\right\}}

\newcommand{\abs}[1]{\left|#1\right|}

\newcommand{\qq}{\qquad}
\newcommand{\ds}{\displaystyle}

\newcommand{\wt}[1]{\widetilde{#1}}

\newcommand{\at}[2]{\genfrac{}{}{0pt}{}{#1}{#2}}

\newcommand{\rmd}{\textrm{d}}

\newcommand{\sn}{\textrm{sn}}

\numberwithin{equation}{section}

\hoffset= - 1.0in
\begin{document}

\title{Modular properties of 6d (DELL) systems}
\author{{\bf G. Aminov$^{a,}$}\footnote{aminov@itep.ru}, {\bf A. Mironov$^{b,a,c,d,}$}\footnote{mironov@itep.ru; mironov@lpi.ru}  \ and  {\bf A. Morozov$^{a,c,d,}$}\thanks{morozov@itep.ru}}
\date{ }

\maketitle

\vspace{-6.0cm}

\begin{center}
\hfill FIAN/TD-22/17\\
\hfill IITP/TH-15/17\\
\hfill ITEP/TH-24/17
\end{center}

\vspace{4.2cm}

\begin{center}
$^a$ {\small {\it ITEP, Moscow 117218, Russia}}\\
$^b$ {\small {\it Lebedev Physics Institute, Moscow 119991, Russia}}\\
$^c$ {\small {\it National Research Nuclear University MEPhI, Moscow 115409, Russia }}\\
$^d$ {\small {\it Institute for Information Transmission Problems, Moscow 127994, Russia}}
\end{center}

\vspace{1cm}

\abstract

If super-Yang-Mills theory possesses the exact conformal invariance,
there is an additional modular invariance under the change of
the complex bare charge $\tau = \frac{\theta}{2\pi}+ \frac{4\pi\imath}{g^2}
\longrightarrow -\frac{1}{\tau}$.
The low-energy Seiberg-Witten prepotential ${\cal F}(a)$, however,
is not explicitly invariant, because the flat moduli also change
$a \longrightarrow a_D = \partial{\cal F}/\partial a$.
In result, the prepotential is not a modular form and
depends also on the anomalous Eisenstein series $E_2$.
This dependence is usually described by the universal
MNW modular anomaly equation.
We demonstrate that, in the $6d$ $SU(N)$ theory with {\it two} independent
modular parameters $\tau$ and $\hat \tau$, the modular anomaly equation changes,
because the modular transform of $\tau$ is accompanied by
an ($N$-dependent!) shift of $\hat\tau$ and vice versa.
This is a new peculiarity of double-elliptic systems,
which deserves further investigation.

\section{Introduction}

Lifting the Seiberg-Witten-Nekrasov \cite{SW1,SW2,GKMMM'95,DW'96,LMNS1,LMNS2,LMNS3,LMNS4,N'04,NO'03,NS'09,MM2'2009,MM3'2009}
theory to the level of $6d$ SYM is now attracting
increasing interest \cite{IKY'2015,Nieri'2015,MMZ1'16,MMZ2'16,MMZ3'16,KP'16}.
One of the research directions here is the interpretation of  the
corresponding Nekrasov functions in terms of the representation theory
of DIM algebras \cite{AFS'2012,NagITEP1'16} and network models \cite{MZ'16,MMZ3'16},
which generalize the Dotsenko-Fateev (conformal matrix model \cite{CFTmamo'91,CFTmamo1'93,CFTmamo2'93,CFTmamo1'95,CFTmamo2'95,CFTmamo3'95}) realization
of conformal blocks, manifest an explicit spectral duality
\cite{MTV,MTV2,BPTY,MMZZ13,MMRZZ13,MMRZZ34,MMZ1'16,MMZ2'16} and satisfy the Virasoro/W-constraints in the
form of the $qq$-character equations \cite{KP'15,MMZ3'16,NagITEP1'16,BFMZZ'16,BFHMZ'17}.
Another direction is study of the underlying integrable systems,
where the main unknown ingredient is the double-elliptic (DELL)
generalization \cite{BMMM'2000,MM,MM2,AMMZ,ABMMZ,AMM'16} of the Calogero-Ruijsenaars model \cite{Calogero'75,Calogero'76,Moser'75,OP'81,Ruijs,Ruj'95,FGNR,GM}.
The both approaches are currently technically involved and not yet
very well related.
In this paper, we demonstrate that, despite the complexity of the subject,
one can already formulate very clear and elegant statements
extracted from a series of pretty sophisticated and tedious calculations.
This is a sign that the whole 6d/DIM/DELL story will finally
acquire a simple and transparent form suitable for a text-book level presentation.

$\mathcal{N}=2$ supersymmetric gauge theories can be studied in the string theory framework, which provides a transparent description for the Coulomb branch of such models. Since we are interested in the low energy effective actions and the corresponding integrable systems, it is useful to formulate the gauge theories under consideration as the quantum field theories derived from various configurations of branes in the superstring and $M$ theory. Let us start with the gauge theories in four dimensions and recall their description via $M$ theory introduced by E. Witten in \cite{Witten'97}, which was a continuation of a series of previous studies in \cite{HW'97,BHOOY'97,BHOY'97,EGK'97}. According to \cite{Witten'97}, a wide class of $4d$ gauge theories can be obtained by considering D4 branes extended between NS5 branes in Type IIA superstring theory on $\mathbb{R}^{10}$ with coordinates $x^0,x^1,\dots,x^9$. The worldvolumes of NS5 branes are six dimensional with coordinates $x^0,x^1,\dots,x^5$ and the  worldvolumes of D4 branes are five dimensional with coordinates $x^0,x^1,x^2,x^3,x^6$. One can locate the NS5 branes at $x^7=x^8=x^9=0$ and, in the classical approximation, at some fixed values of $x^6$, while the D4 branes are finite in the $x^6$ direction and terminate on the NS5 branes. Following \cite{Witten'97}, we introduce a complex variable $v=x^4+\imath x^5$ and, classically, every D4 brane is located at a definite value of $v$. Such brane configurations can be illustrated by the following picture with vertical and horizontal directions being $v$ and $x^6$ correspondingly:
\eq{\includegraphics[height=6cm]{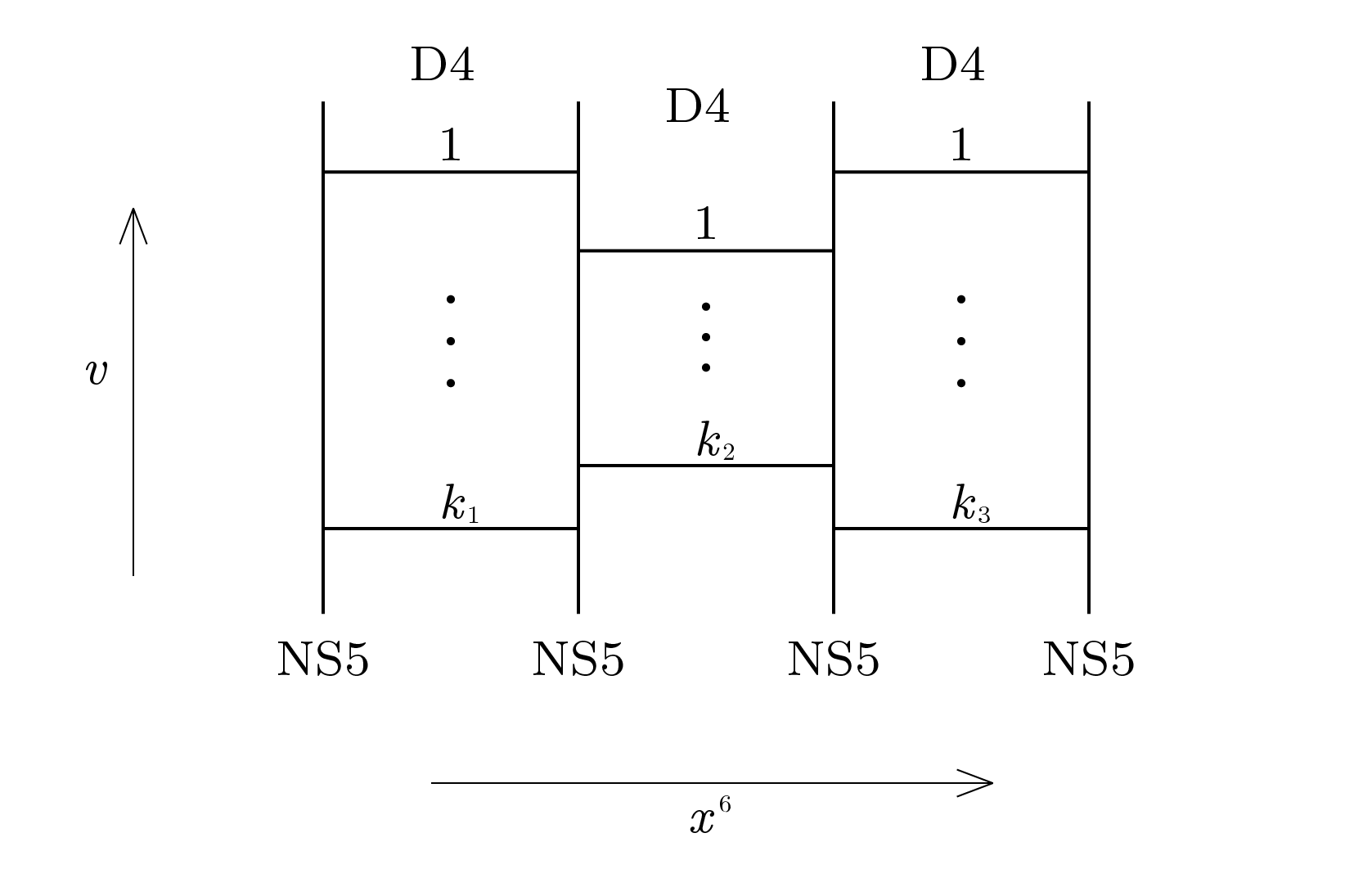}}
If one has $n+1$ fivebranes labeled by $\alpha=0,\dots,n$, and $k_{\alpha}$ fourbranes attached to the $\brc{\alpha-1}$-th and $\alpha$-th fivebranes, the gauge group of the four-dimensional theory is $\prod_{\alpha=1}^n SU\brc{k_{\alpha}}$.
The positions of the fourbranes $a_{i,\alpha}$, $i=1,\dots,k_{\alpha}$ correspond to the Coulomb moduli of the gauge theory. The coupling constant $g_{\alpha}$ of the $SU\brc{k_{\alpha}}$ gauge group is given by
\eqlb{eq:gcouplings_a}{\frac1{g_{\alpha}^2}=\frac{x^6_{\alpha}-x^6_{\alpha-1}}{\lambda},}
where $x^6_{\alpha}$ is the position of the $\alpha^{\textrm{th}}$ fivebrane in the $x^6$ direction and $\lambda$ is the string coupling constant. In fact, the fivebranes do not really have any definite values of $x^6$ as the classical brane picture suggests. The position $x^6_{\alpha}$ is determined as a function of $v$ by minimizing the total fivebrane worldvolume. Thus, $g_{\alpha}$ is also a function of $v$ and $g_{\alpha}\brc{v}$ can be interpreted as the effective coupling of the $SU\brc{k_{\alpha}}$ theory at mass $\abs{v}$. To include the effective theta angle $\theta_{\alpha}$ of the $SU\brc{k_{\alpha}}$ gauge theory, one has to lift Type IIA superstring theory to the $M$ theory on the $\mathbb{R}^{10}\times S^1$. The eleventh dimension $x^{10}$ in $M$ theory is periodic with period $2 \pi R_{10}$. The theta angle $\theta_{\alpha}$ is determined by the separation in the $x^{10}$ direction between the $\brc{\alpha-1}^{\textrm{th}}$ and $\alpha^{\textrm{th}}$ fivebranes:
\eq{\theta_{\alpha}=\frac{x^{10}_{\alpha}-x^{10}_{\alpha-1}}{R_{10}}.}
Then the complexified coupling constant is
\eq{-\imath \tau_{\alpha}=\frac{4\pi}{g_{\alpha}^2}-\frac{\imath\,\theta_{\alpha}}{2\pi}.}

The brane configuration is different in M theory. In general, the Type IIA fivebrane on $\mathbb{R}^{10}$ corresponds to the M5 brane on $\mathbb{R}^{10}\times S^1$ located at a point in $S^1$, and the Type IIA fourbrane corresponds to the M5 brane that is wrapped over the $S^1$. As it was described in \cite{Witten'97}, the Type IIA configuration of the NS5 branes joined by the D4 branes corresponds in M theory to a single M5 brane with a more complicated world history. The worldvolume of this M5 brane is $\mathbb{R}^{4}\times \Gamma$, where $\mathbb{R}^{4}$ is parameterized by the first four coordinates $x^0,x^1,x^2,x^3$ and $\Gamma$ is a two-dimensional surface in $\mathbb{R}^{3}\times S^1$ parameterized by $x^4,x^5,x^6,x^{10}$. If we provide $\mathbb{R}^{3}\times S^1$ with the complex structure with holomorphic variables $v=x^4 +\imath x^5$ and $s=x^6 +\imath x^{10}$, then, due to the $\mathcal{N}=2$ supersymmetry, $\Gamma$ is a complex Riemann surface.
This surface plays a great role in connecting M theory with the theory of integrable systems \cite{GGM1'98,MMM'98,GGM2'98,GM}. In particular, the low energy effective action of the $\mathcal{N}=2$ gauge theory can be determined by an integrable Hamiltonian system \cite{GKMMM'95,MW'96,DW'96} with the spectral curve given by $\Gamma$, which is usually called the Seiberg-Witten curve.

In this paper, we use methods from the theory of integrable systems to study some particular curves $\Gamma$ and the corresponding low energy effective actions. We focus on a special case of systems with $x^6$ direction compactified onto a circle. This case describes theories with adjoint matter hypermultiplets, their bare masses $m_{\alpha}$ being given by differences between the average positions
in the $v$ plane of the fourbranes to the left and right of the $\alpha^{\textrm{th}}$ fivebranes:
\eq{m_{\alpha}=\frac1{k_{\alpha}}\sum_i a_{i,\alpha}-\frac1{k_{\alpha+1}}\sum_j a_{j,\alpha+1}.}
Besides, the numbers of D4 branes $k_\alpha$ are all coincide and the gauge group is $U(1)\times SU(k)^n$.
Various brane configurations provide us with gauge theories of this type in different dimensions. From the M theory point of view, there is a natural set of gauge theories in dimensions $4$, $5$ and $6$. First, consider the $4d$ case and the following brane configuration in Type IIA theory on $\mathbb{R}^{9}\times S^1$:
\eqlb{fig:2}{\includegraphics[height=6.5cm]{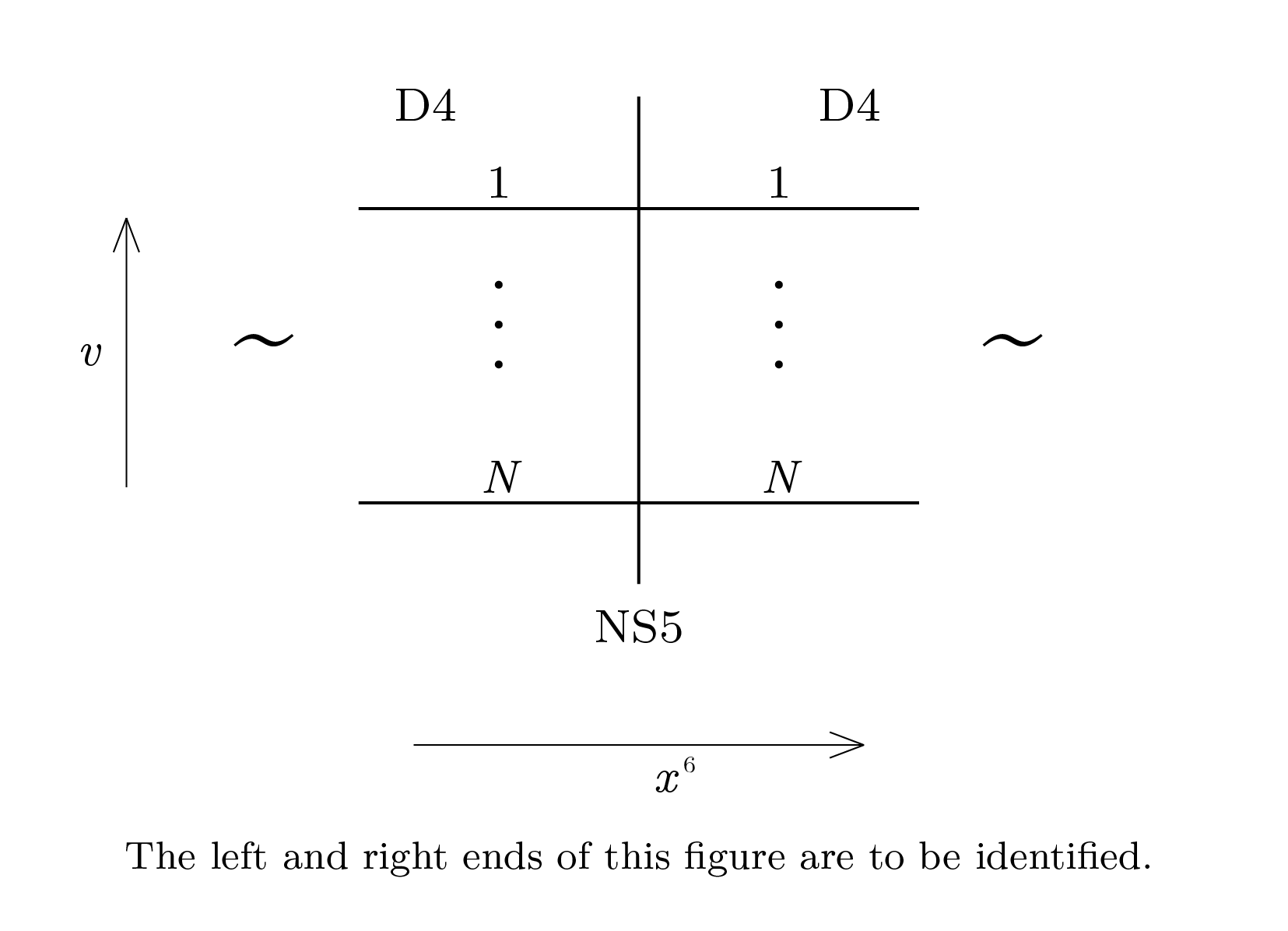}}
where there is one NS5 brane and $N$ D4 branes wrapped around a circle in the $x^6$ direction, i.e. the gauge group is $U\brc{1}\times SU\brc{N}$.
In fact, the particular configuration depicted in figure \ref{fig:2} corresponds to the $\mathcal{N}=4$ theory with gauge group $U\brc{k}$, because the hypermultiplet bare mass is zero. This is due to the simple choice of the spacetime, which, in coordinates $x^6$ and $v=x^4+\imath x^5$, is just $S^1\times\mathbb{C}$. Thus, each D4 brane is ending at the same point to the left and right of the NS5 brane, resulting in zero difference between the average positions of the fourbranes on two sides of the fivebrane. To introduce a non-zero hypermultiplet bare mass and to break the $\mathcal{N}=4$ supersymmetry down to $\mathcal{N}=2$, one needs to replace $S^1\times\mathbb{C}$ part of the spacetime by a certain $\mathbb{C}$ bundle over $S^1$. The procedure introduced in \cite{Witten'97} is to start with $x^6$ and $v$ as coordinates on $\mathbb{R}\times\mathbb{C}$ and divide by the following symmetry:
\eqlb{eq:IIAsym}{
\arraycolsep=2pt
\begin{array}{rcl}
x^6 & \rightarrow & x^6 + 2 \pi L,\\
v   & \rightarrow & v + m,
\end{array}}
where an arbitrary complex constant $m$ defines the hypermultiplet bare mass and the corresponding type IIA brane configuration is
\eqlb{pic:figure3}{\includegraphics[height=6cm]{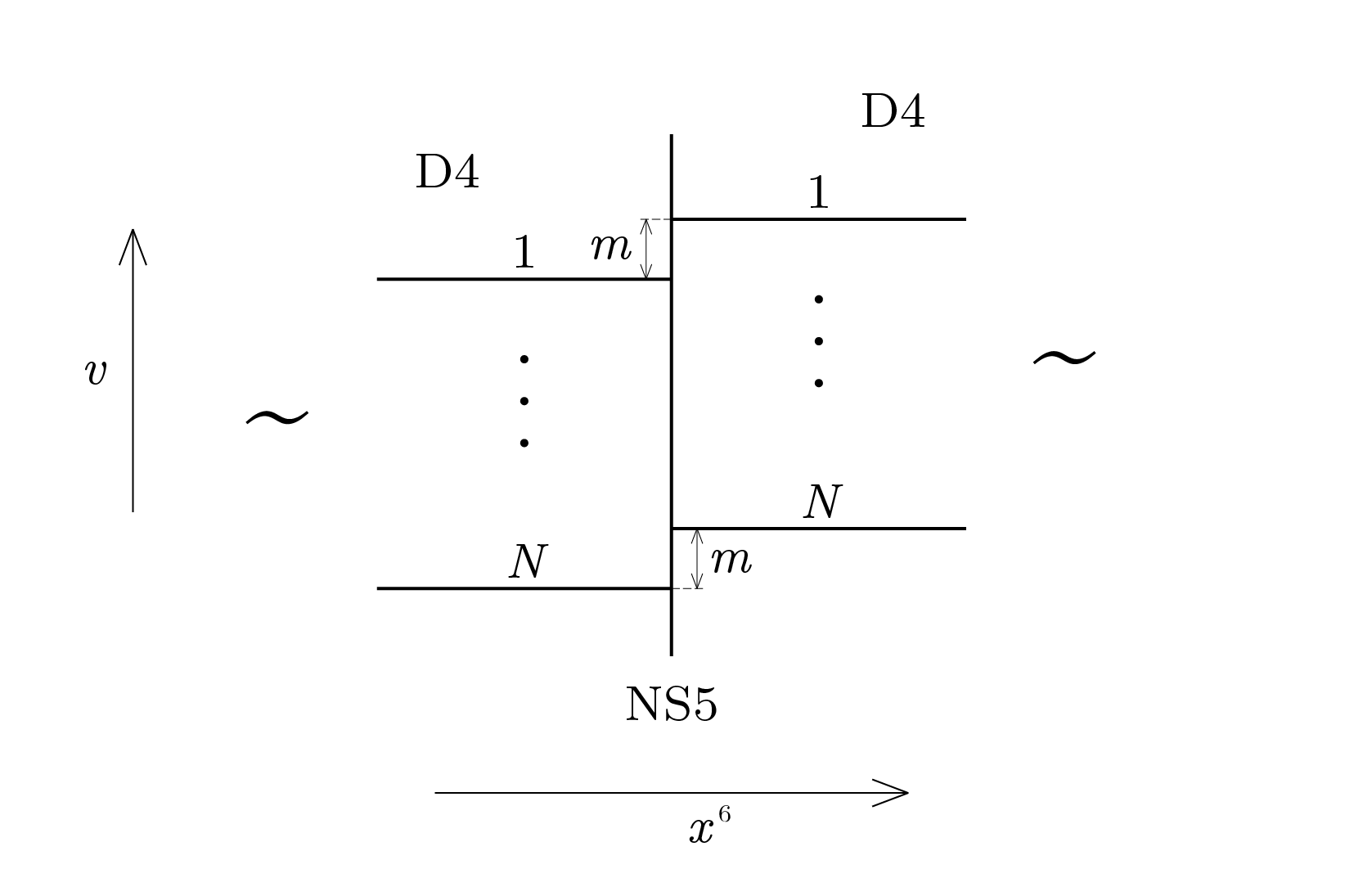}}
Now, upon going around the $x^6$ circle, one comes back with a shifted value of $v$. The M theory uplift of this model also requires some particular choice of the spacetime. To get a non-zero theta angle, one divides $\mathbb{R}\times S^1\times\mathbb{C}$ part of the spacetime with coordinates $x^6$, $x^{10}$, and $v$ by the combined symmetry
\eqlb{eq:Msym}{
\arraycolsep=2pt
\begin{array}{rcl}
x^6  & \rightarrow & x^6 + 2 \pi L,\\
x^{10} & \rightarrow & x^{10} + \theta,\\
v    & \rightarrow & v + m,
\end{array}}
where $\theta$ defines the effective theta angle and $x^{10}$ is still periodic with period $2 \pi R_{10}$. The quotient of the $s$ plane by these equivalences, i.e. of the $\mathbb{R}\times S^1$ part of the space is a complex Riemann surface $\Sigma$ of genus one with modulus $\tau$ giving the complexified coupling constant of the theory. The resulting quotient of the whole $\mathbb{R}\times S^1\times\mathbb{C}$ by (\ref{eq:Msym}) is a complex manifold $X_m$, which can be regarded as a $\mathbb{C}$ bundle over $\Sigma$. The type IIA brane configuration (\ref{pic:figure3}) in terms of M theory is described by a single M5 brane, which propagates in $X_m$. The worldvolume of this fivebrane is given by $\mathbb{R}^{4}\times \Gamma$, where $\Gamma$ is a two-dimensional Riemann surface in $X_m$. An important part of the $X_m$ structure is the map $X_m\rightarrow \Sigma$ provided by forgetting $\mathbb{C}$. Under this map, the curve $\Gamma\subset X_m$ maps to $\Sigma$, thus giving an interpretation of $\Gamma$ as an $N$-sheeted covering of the base torus $\Sigma$. From the viewpoint of integrable systems, $\Gamma$ corresponds to the spectral curve $\Gamma^{\textrm{CM}}$ of the elliptic Calogero-Moser model \cite{DW'96,IM'1996,IM'1997} known to have the same geometrical description \cite{HPh'98} (generalization to the case of more than two NS5 branes leads to the spin Calogero model, see \cite{MMZ2'16}). To avoid uncertainties in the notation, from now on, we denote the curve $\Gamma$ of the $4d$ theory under consideration by $\Gamma^{\textrm{CM}}$.

Before going to the $5d$ and $6d$ cases, we briefly review some basic properties of the curve $\Gamma^{\textrm{CM}}$ and the corresponding low energy effective action. Theories resulting from the brane configurations described above, with the $x^6$ direction compactified onto a circle, are known to be conformally invariant \cite{Witten'97}. The duality group of the four-dimensional model is  $SL\brc{2,\mathbb{Z}}$. In other words, the curve $\Gamma^{\textrm{CM}}$ is invariant under the modular transformations $\tau\rightarrow\tau+1$ and $\tau\rightarrow-1/\tau$. The low energy effective action is not invariant, but has very distinctive properties under the action of the duality group. These properties can be understood by describing the low energy effective action in terms of the Seiberg-Witten prepotential $\mathcal{F}^{\textrm{CM}}$, whose second derivatives with respect to the Coulomb moduli $a_i$ give the period matrix $T^{\textrm{CM}}$ of the complex Riemann surface $\Gamma^{\textrm{CM}}$. Using this connection between the prepotential $\mathcal{F}^{\textrm{CM}}$ and the curve $\Gamma^{\textrm{CM}}$, the modular anomaly equation describing the dependence of $\mathcal{F}^{\textrm{CM}}$ on the second Eisenstein series $E_2\brc{\tau}$ was derived by J. Minahan, D. Nemeschansky and N. Warner in \cite{MNW}.
This equation has an elegant form
\eqlb{eq:Intro_Mod4d}{
\frac{6}{\pi\imath}\frac{\partial \mathcal{F}^{\textrm{CM}}}{\partial E_2}=
-\frac{1}{2}\sum_{i=1}^{N}\brc{\frac{\partial \mathcal{F}^{\textrm{CM}}}{\partial a_i}-\tau\,a_i}^2}
and is equivalent to the holomorphic anomaly equation \cite{HIKLV'13} in the limit of $\epsilon_1,\epsilon_2\rightarrow 0$. Brane configuration also provides valuable insights into the dependence of the low energy effective action on the Coulomb moduli. Since the $U\brc{1}$ factor decouples from the $SU\brc{N}$ part of the theory, the period matrix $T^{\textrm{CM}}$ depends only on the differences $\brc{a_i-a_j}$. In Type IIA theory, the Coulomb moduli $a_i$ describe the positions of the fourbranes in the $v$ plane, and these fourbranes are all identical. Therefore, the curve $\Gamma^{\textrm{CM}}$ in M theory is invariant under permutations of the moduli $a_i$, and the period matrix is a symmetric function of the differences $\brc{a_i-a_j}$. The same is true for the perturbative and instanton parts of the prepotential $\mathcal{F}^{\textrm{CM}}$. Another basic aspect of the theory is its behavior at particular values of the bare mass $m$. As it was mentioned earlier,  $\mathcal{N}=2$ theory with gauge group  $U\brc{1}\times SU\brc{N}$ becomes $\mathcal{N}=4$ theory with gauge group $U\brc{k}$ at $m=0$. Thus, the prepotential $\mathcal{F}^{\textrm{CM}}$ at $m=0$ is
\eq{\left.\mathcal{F}^{\textrm{CM}}\right|_{m=0}=\frac{\tau}2\sum_{i=1}^{N} a^2_i,}
which is associated with the classical part of the prepotential.
To obtain the pure gauge limit of the  $\mathcal{N}=2$ theory, one should bring the value of $m$ and $\imath\,\tau$ to infinity in a consistent way (double scaling limit) \cite{Ino'89,IM'1996,HPh'98,MNW}:
\eqlb{eq:4dlim_intro}{m\rightarrow\infty,\quad \textrm{Im}\,\tau \rightarrow+\infty,\quad
m^{2N}\exp\brc{2\pi\imath\,\tau}\rightarrow\brc{-1}^N\Lambda^{2N},}
so that the resulting cutoff $\Lambda$ is finite. From the M theory point of view, this limit of infinite mass in the four-dimensional theory is accompanied with the decompactification of the $x^6$ direction.

Roughly speaking, the $5d$ and $6d$ theories can be obtained by successively compactifying the $x^4$ and $x^5$ directions in M theory. To get the proper gauge theory description, one should start with Type IIA superstring theory and perform the $T$-duality transformation that turns Type IIA theory into Type IIB. In this way, the five-dimensional gauge theory can be described in terms of the Type IIB D5 and NS5 branes, which form a Type IIB $\brc{p,q}$-brane web \cite{KKV'97,GGM2'98,KMV'98,Kol'99,AHK'98,KR'98}. For our purposes of studying the Seiberg-Witten curves and the low energy effective actions of the $5d$ and $6d$ theories, it is sufficient to use the earlier described configuration of the single M5 brane and further compactify the $x^4$ and $x^5$ directions. In particular, $5d$ SYM theory with one compactified Kaluza-Klein dimension and the adjoint matter hypermultiplet \cite{BMMM'99,BMMM'1999,Nek5} corresponds to the brane configuration with $x^4$ direction compactified onto a circle of radius $R_4=\beta^{-1}/2$. The part of the spacetime with coordinates $x^6$, $x^{10}$, and $v=x^4+\imath x^5$ is divided by the symmetry
\eqlb{eq:Msym_5d}{
\arraycolsep=2pt
\begin{array}{rcl}
x^6  & \rightarrow & x^6 + 2 \pi L,\\
x^{10} & \rightarrow & x^{10} + \theta,\\
v    & \rightarrow & v + \beta^{-1}\epsilon,
\end{array}}
providing a complex manifold $X_\epsilon$, where $\epsilon$ defines the hypermultiplet bare mass.
The worldvolume of M5 brane in the $5d$ case is $\mathbb{R}^{4}\times \Gamma^{\textrm{RS}}$ with $\Gamma^{\textrm{RS}}\subset X_\epsilon$ and the curve $\Gamma^{\textrm{RS}}$ is equivalent to the spectral curve of the elliptic Ruijsenaars system \cite{BMMM'1999}. The compactification of $x^4$ direction affects the low energy effective action and the curve $\Gamma^{\textrm{RS}}$ in a very manifest way. Since the Coulomb moduli $a_i$ take values in the $v$ plane with the periodic real coordinate $x^4$, the curve should be invariant under the shifts $a_i\rightarrow a_i +\pi\,\beta^{-1}$. Thus, the period matrix $T^{\textrm{RS}}$ can be represented as a symmetric function of $\sin\brc{\beta\, a_{ij}}^2$ with $a_{ij}\equiv a_i - a_j$. According to (\ref{eq:Msym_5d}), the mass parameter $\epsilon$ describes the shift in the $v$ plane, and there should be another symmetry of the curve $\Gamma^{\textrm{RS}}$, that is, $\epsilon\rightarrow\epsilon +\pi$. The $5d$ theory under consideration is conformally invariant and the duality group is  $SL\brc{2,\mathbb{Z}}$. As it was established in several works \cite{AMM'16,KimNahm'17}, the Seiberg-Witten prepotential $\mathcal{F}^{\textrm{RS}}$ admits the same modular anomaly equation (\ref{eq:Intro_Mod4d}) as in the $4d$ case. Also, at $\epsilon=0$, the  $\mathcal{N}=2$ supersymmetry becomes $\mathcal{N}=4$ and
\eq{\left.\mathcal{F}^{\textrm{RS}}\right|_{\epsilon=0}=\frac{\tau}2\sum_{i=1}^{N} a^2_i.}
The pure gauge limit of the $5d$ theory, however, is different. The curve is invariant under $\epsilon\rightarrow\epsilon +\pi$, and $T^{\textrm{RS}}$ depends on $\epsilon$ only through $\brc{\sin\epsilon}^2$. This results in the following definition of the $5d$ cutoff $\wt\Lambda$:
\eqlb{eq:5dlim_intro}{\textrm{Im}\,\epsilon\rightarrow\infty,\quad \textrm{Im}\,\tau \rightarrow+\infty,\quad
\brc{\sin\epsilon}^{2N} \exp\brc{2\pi\imath\,\tau}\rightarrow\brc{-1}^N\wt\Lambda^{2N}.}
Again, the limit of infinite mass in the five-dimensional theory is accompanied with the decompactification of the $x^6$ direction.

The most general system that can be obtained in the present setup is the $6d$ SYM theory with two compactified Kaluza-Klein dimensions and the adjoint matter hypermultiplet. The corresponding brane configuration is a single M5 brane in a spacetime, where the $v$ plane is compactified to a torus $S^1\times S^1 = \mathbb{T}^2$ with modulus $\hat\tau=\imath\, R_5/ R_4$, and $R_5$ is the radius of the $x^5$ direction. The $\mathbb{R}\times S^1\times \mathbb{T}^2$ part of the spacetime with coordinates $x^6$, $x^{10}$, and $v$ is divided by the symmetry (\ref{eq:Msym_5d}), and the resulting quotient is a complex manifold $X_{\brc{\epsilon,\hat\tau}}$, which can be regarded as a $\mathbb{T}^2$ bundle over $\Sigma$. The two-dimensional Riemann surface $\Gamma^{\textrm{Dell}}\subset X_{\brc{\epsilon,\hat\tau}}$, which is a part of the M5 brane worldvolume $\mathbb{R}^{4}\times \Gamma^{\textrm{Dell}}$, corresponds to the spectral curve of the double-elliptic integrable system \cite{BMMM'2000,MM,MM2,Braden2003} of $N$ interacting particles. The term {\it double-elliptic} reflects the fact that there are two elliptic curves, $\Sigma$ and $\mathbb{T}^2$ with moduli $\tau$ and $\hat\tau$ correspondingly. Since under the map $X_{\brc{\epsilon,\hat\tau}}\rightarrow\Sigma$ the curve $\Gamma^{\textrm{Dell}}$ maps to $\Sigma$, we consider $\Gamma^{\textrm{Dell}}$ as an $N$-sheeted covering of the base torus $\Sigma$. This system can be also described with the help of Type IIB theory, and the relevant $\brc{p,q}$-brane web was introduced recently in \cite{MMZ2'16}.
Similar to the $5d$ case, the compactness of the forth and fifth spacetime dimensions can be used to describe some basic properties of the low energy effective action. The Coulomb moduli $a_i$ now take values in the torus $\mathbb{T}^2$, which means that there is an additional symmetry $a_i\rightarrow a_i+\pi\,\beta^{-1}\hat\tau$ of the curve $\Gamma^{\textrm{Dell}}$. Thus, the period matrix $T^{\textrm{Dell}}$ should depend on the differences $\brc{a_i-a_j}$ through an elliptic function.
The most common way to obtain such functions is to consider the second logarithmic derivatives of the Riemann theta function. In this paper, we use the function $\sigma\brc{z|\,\hat\tau}$ defined as
\eqlb{eq:def_sigma}{\frac1{\sigma\brc{z|\,\hat\tau}^2}\equiv -\partial_{z}^2 \log\theta_{11}\brc{\pi^{-1}z|\,\hat\tau},}
where $\theta_{11}\brc{\pi^{-1}z|\,\hat\tau}$ is the usual notation for the Riemann theta function with characteristics $\brc{1/2, 1/2}$:
\eq{\theta_{11}\brc{\pi^{-1}z|\,\hat\tau}=\sum_{n\in\mathbb{Z}} \exp\brc{\pi\imath \brc{n+1/2}^2 \hat\tau + 2\pi \imath \brc{n+1/2}\brc{z+1/2}}.}
For small $z$, (\ref{eq:def_sigma}) can be rewritten with the help of the Eisenstein series $\bfi{E_{2k}}$ and of the Riemann zeta function $\zeta\brc{k}$:
\eqlb{eq:defSigma}{\frac1{\sigma\brc{z|\,\hat\tau}^2}= \frac1{z^2} + 2\sum_{k=1}^{+\infty} \frac{\zeta\brc{2k}}{\pi^{2k}}
\brc{2k-1} E_{2k}\brc{\hat\tau}\,z^{2k-2}.}
One could expect that the dependence of the period matrix on the mass parameter $\epsilon$ is also through an elliptic function. However,  the curve $\Gamma^{\textrm{Dell}}$ is not invariant under the shift $\epsilon\rightarrow \epsilon +\pi\,\hat\tau$ alone.
It turns out that the shift of the mass parameter is accompanied with the shift of the first elliptic parameter $\tau$, and the actual symmetry of $\Gamma^{\textrm{Dell}}$ is
\eqlb{eq:esym6d}{\epsilon\rightarrow\epsilon+\pi\,\hat\tau, \quad \tau\rightarrow\tau + N \brc{\hat\tau+ 2\pi^{-1}\epsilon}.}
Since this symmetry is observed in the low energy limit of the theory, it probably has more involved structure in the superstring and M theory. Nonetheless, the following elementary interpretation can be suggested. In Type IIA theory, $\epsilon$ describes the distance on $\mathbb{T}^2$ between the two ends of a D4 brane. The brane configuration in  $S^1\times \mathbb{T}^2$ part of the spacetime with one D4 brane can be represented by the following embedding into the three-dimensional space:
\eq{\includegraphics[height=4cm]{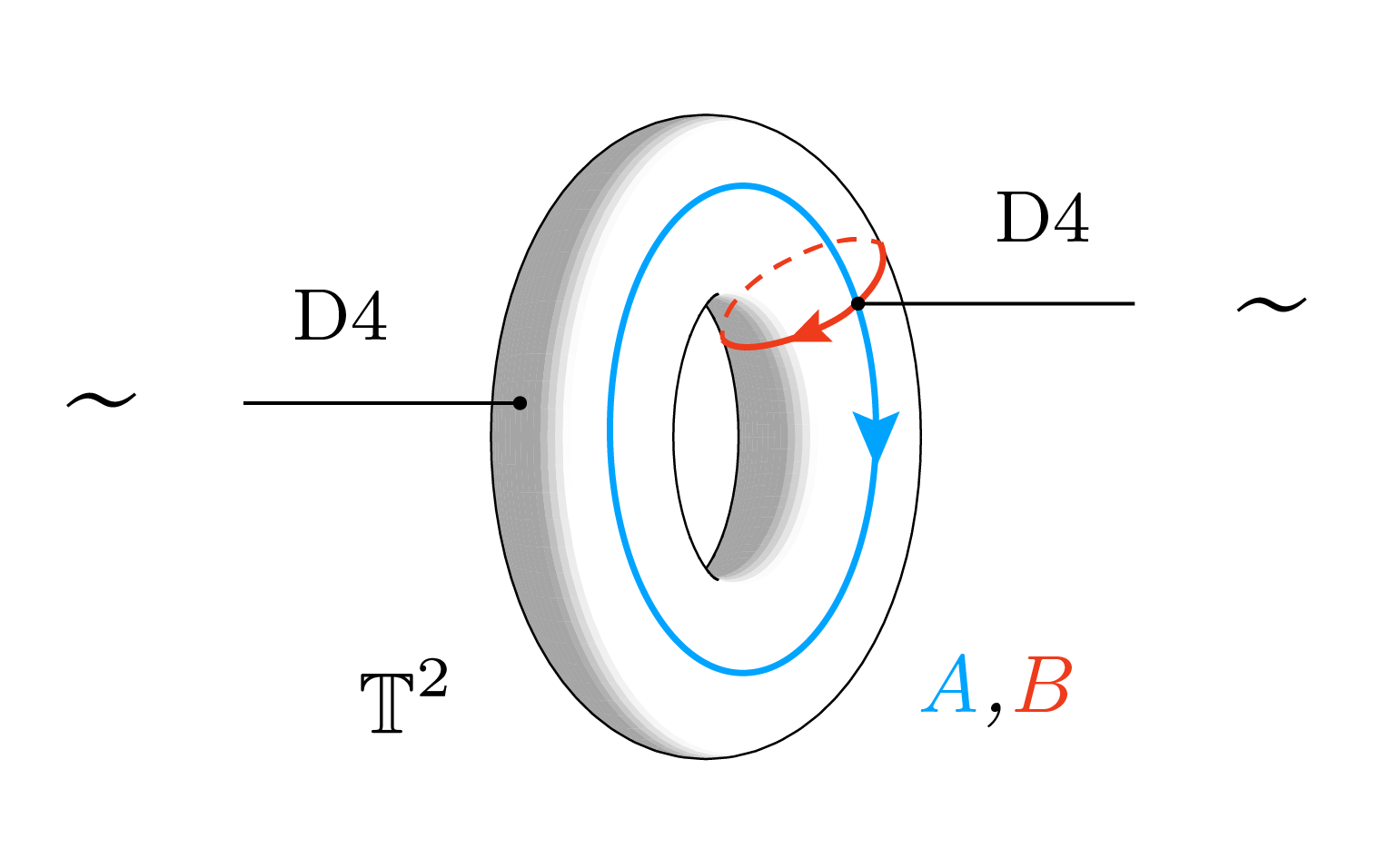}}
The $A$ and $B$ cycles on $\mathbb{T}^2$ correspond to the compactified $x^4$ and $x^5$ directions respectively. Upon moving one end of the D4 brane all the way around the $x^4$ direction, the line representing this D4 brane in the $3d$ embedding goes around the $A$ cycle, and we get the same configuration we started with. This describes the symmetry $\epsilon\rightarrow\epsilon +\pi$. When we move one end of the fourbrane all the way around the $x^5$ direction, the line in the $3d$ embedding wraps around the $B$ cycle. This could be interpreted as some effective extension of the fourbrane length or the radius of the $x^6$ direction. In M theory, this D4 brane becomes a part of a single M5 brane and its wrapping around $x^5$ direction could be interpreted as some effective shift of the first elliptic parameter $\tau$. The above interpretation is based on the particular form of the $3d$ embedding and does not explain the exact value of the shift in $\tau$. As a result of the symmetry (\ref{eq:esym6d}), the period matrix $T^{\textrm{Dell}}$ depends on the mass parameter $\epsilon$ not only through the elliptic function $\sigma\brc{\epsilon|\,\hat\tau}$, but also through the Riemann theta function $\theta_{11}\brc{\pi^{-1}\epsilon|\,\hat\tau}$, which will be seen later in formulas (\ref{eq:pert_N}), (\ref{eq:c1_q1Exp}), and (\ref{eq:C100_q1Exp}).
After compactifying the $x^5$ direction, the theory remains conformally invariant, but the duality group changes. The obvious reason is that now one has two elliptic curves and two duality groups describing the modular transformations of two elliptic parameters $\tau$ and $\hat\tau$.
An essential observation made in this paper is that the duality group is not just a product $SL\brc{2,\mathbb{Z}}\times SL\brc{2,\mathbb{Z}}$. Instead, the modular transformation of one of the elliptic parameters is accompanied by the shift of the other and, for generic values of the parameters of the theory, this shift is not even an element of the group $SL\brc{2,\mathbb{Z}}$ .
The four generators of this duality group are
\eqlb{eq:sym_gen}{\begin{array}{lll}
\left.1\right) & \tau\rightarrow\tau+1, & \hat\tau\rightarrow\hat\tau,\\
\left.2\right) & \tau\rightarrow-1/\tau, & \hat\tau\rightarrow\hat\tau- \pi^{-2} N\epsilon^2\tau^{-1},\\
\left.3\right) & \hat\tau\rightarrow\hat\tau+1, & \tau\rightarrow\tau,\\
\left.4\right) & \hat\tau\rightarrow-1/\hat\tau, & \tau\rightarrow\tau- \pi^{-2} N\epsilon^2\hat\tau^{-1}.\\
\end{array}}
The actions of the second and the forth generators from (\ref{eq:sym_gen}) on the Seiberg-Witten prepotential $\mathcal{F}^{\textrm{Dell}}$ can be described by two modular anomaly equations. The first equation is a generalization of the four-dimensional MNW modular anomaly equation and has one additional term, the derivative of the prepotential with respect to the second elliptic parameter $\hat\tau$:
\eqlb{eq:Intro_Mod6d}{\boxed{
\frac{6}{\pi\imath}\frac{\partial \mathcal{F}^{\textrm{Dell}}}{\partial E_2} - \frac{N\epsilon^2}{\pi^2} \frac{\partial \mathcal{F}^{\textrm{Dell}}}{\partial \hat\tau}=
-\frac{1}{2}\sum_{i=1}^{N}\brc{\frac{\partial \mathcal{F}^{\textrm{Dell}}}{\partial a_i}-\tau\,a_i}^2.}}
The second modular anomaly equation is
\eqlb{eq:Intro_Mod6d_2}{\boxed{
\frac{6}{\pi\imath}\frac{\partial \mathcal{F}^{\textrm{Dell}}}{\partial \hat E_2} - \frac{N\,\epsilon^2}{\pi^2} \frac{\partial \mathcal{F}^{\textrm{Dell}}}{\partial \tau}=
-\frac{\epsilon^2}{2\,\pi^2}\brc{\sum_{i=1}^{N}a_i}^2}}
with the notation $\hat E_2\equiv E_2\brc{\hat\tau}$. At this point, one can see that, in the present setup, the low energy effective action is not invariant under the simple permutation of the two elliptic parameters $\tau$ and $\hat\tau$. This is because we started with Type IIA theory and, within the obtained formulation of M theory, the two tori $\Sigma$ and $\mathbb{T}^2$ are not exactly equivalent. These tori could become equivalent after a series of $T$-dualities and appropriate changes of the spacetime.
We expect that the $6d$ modular anomaly equations can be lifted to the level of Nekrasov functions, as it was done for the $4d$ case in \cite{BFL1,BFL2,BFL3,BFL4} and to the level of $2d$ conformal field theories in \cite{GMM,Nem}. Note that, in the recent paper by S. Kim and J. Nahmgoong,  \cite{KimNahm'17}, the $S$-duality in $6d$ $\brc{2,0}$ theory was studied. From the point of view of SYM theories, the partition function considered in \cite{KimNahm'17} corresponds to the Nekrasov instanton partition function of the $5d$ SYM theory with the adjoint matter hypermultiplet. One of the results described in \cite{KimNahm'17} is that the $5d$ prepotential admits the same modular anomaly equation as the $4d$ one, in accordance with what was stated in \cite{AMM'16}.

One more topic we are going to discuss in this paper is the behavior of the theory at particular values of the bare mass $\epsilon$. At $\epsilon=0$, the theory becomes $\mathcal{N}=4$ supersymmetric theory, and the prepotential is
\eqlb{eq:dell_N=4}{\left.\mathcal{F}^{\textrm{Dell}}\right|_{\epsilon=0}=\frac{\tau}2\sum_{i=1}^{N} a^2_i.}
Since $\epsilon$ describes the shift in two compact dimensions, there is neither the limit of infinite mass nor the pure gauge limit in the $6d$ case. Yet there is a special point $\epsilon=\epsilon^{\infty}$, at which the elliptic function $\sigma\brc{\epsilon|\,\hat\tau}$ goes to infinity. In fact, an elliptic function must have at least two poles in a fundamental parallelogram, but we will use the single notation $\epsilon^{\infty}$ keeping in mind that $\epsilon^{\infty}$ can take several values.
The exact value of $\epsilon^{\infty}$ depends on the particular choice of elliptic function, and, in our case, it can be described as a solution to the following equation:
\eqlb{eq:einf_intro}{\wp\brc{\epsilon^{\infty}|\,\hat{\tau}}=-\frac{1}{3}E_2\brc{\hat\tau},}
where relation (\ref{eq:appsigmaW}) between the $\sigma$ function and the Weierstrass $\wp$ function was used. By analogy with the $4d$ and $5d$ cases, we consider the limit:
\eqlb{eq:6dlim_intro}{\epsilon\rightarrow\epsilon^{\infty},\quad
\sigma\brc{\epsilon|\,\hat\tau}\rightarrow\infty, \quad
\frac{\theta_{11}\brc{\pi^{-1}\epsilon |\,\hat\tau}^{2N}}
{\pi^{-2N}\, \theta_{11}'\brc{0|\,\hat\tau}^{2N}} \exp\brc{2\pi\imath\,\tau}\rightarrow \brc{-1}^N\hat\Lambda^{2N},}
where the new parameter $\hat\Lambda$ plays the role of the effective cutoff in the prepotential. Since the Riemann theta function $\theta_{11}\brc{\pi^{-1}\epsilon|\,\hat\tau}$ has no poles and is finite at $\epsilon=\epsilon^{\infty}$, there is no need to decompactify the $x^6$ direction and bring the first elliptic parameter $\tau$ to the imaginary infinity. In what follows, we refer to (\ref{eq:6dlim_intro}) as the limit $\sigma\brc{\epsilon|\,\hat\tau}\rightarrow\infty$. Despite all the differences, one can still recover (\ref{eq:5dlim_intro}) and (\ref{eq:4dlim_intro}) from (\ref{eq:6dlim_intro}) by considering the limit of $\textrm{Im}\,\hat\tau\rightarrow+\infty$. Since $\textrm{Im}\,\epsilon^{\infty}$ is proportional to $\textrm{Im}\,\hat\tau$, and (\ref{eq:einf_intro}) implies that $\textrm{Im}\,\epsilon^{\infty}$ is non-zero, $\epsilon^{\infty}$ goes to imaginary infinity in the limit $\textrm{Im}\,\hat\tau\rightarrow+\infty$. Theta functions in (\ref{eq:6dlim_intro}) degenerate into the $\brc{\sin\epsilon}^{2N}$ and to get the finite cutoff one restores the limit $\textrm{Im}\,\tau\rightarrow+\infty$.

The rest of the paper is organized as follows. In section \ref{sec:Dell_Prep}, we introduce the double-elliptic Seiberg-Witten prepotential for $N\geq 2$. In section \ref{sec:First_MA}, we discuss the curve $\Gamma^{\textrm{Dell}}$ and its properties under the modular transformations of the first elliptic parameter $\tau$, which leads to the first modular anomaly equation (\ref{eq:Intro_Mod6d}). In section \ref{sec:Second_MA}, the modular transformations of the second elliptic parameter $\hat\tau$ are studied, and the second modular anomaly equation (\ref{eq:Intro_Mod6d_2}) is derived. The limit $\sigma\brc{\epsilon|\,\hat\tau}\rightarrow\infty$ is described in section \ref{sec:mass_lim}, and the convergency condition is formulated as some nontrivial restriction on the coefficients in the series expansion of the double-elliptic prepotential (\ref{eq:prepDellN}). In section \ref{sec:N=2}, the $N=2$ double-elliptic prepotential is considered. We demonstrate that the first modular anomaly equation along with the convergency condition for the limit $\sigma\brc{\epsilon|\,\hat\tau}\rightarrow\infty$ can be used to calculate this prepotential as a series in the mass parameter $\epsilon$. The second modular anomaly equation also proves to be very efficient in the $N=2$ case, because it reduces the problem of computation of the prepotential to the problem of finding of one single function $\hat c_1 \brc{\epsilon, \tau, \hat\tau}$. In a similar way, we use the first modular anomaly equation and the limit of $\sigma\brc{\epsilon|\,\hat\tau}\rightarrow\infty$ to compute the $N=3$ prepotential in section \ref{sec:N=3}. The results for the $N=3$ case are in complete agreement with the calculations from \cite{AMM'16}, where the involutivity conditions for the double-elliptic Hamiltonians were used to compute the prepotential. For $N\geq 3$, the second modular anomaly equation is not that efficient as in the $N=2$ case. However, it provides nontrivial relations between the coefficients in the series expansion of the double-elliptic prepotential. In both $N=2$ and $N=3$ cases, we evaluate the first few orders in the $q$-expansions,  $q\equiv\exp\brc{2\pi\imath\,\tau}$, of the first nontrivial coefficient $\widehat{C}_{i_1,\dots,i_n}\brc{\epsilon,\tau,\hat\tau}$ in the expansion (\ref{eq:prepDellN}) with $i_1=1$ and $i_2=\dots=i_n=0$. The results given in (\ref{eq:c1_q1Exp}) and (\ref{eq:C100_q1Exp}) clearly manifest the symmetry (\ref{eq:esym6d}) and are consistent with the limit (\ref{eq:6dlim_intro}). Moreover, due to the properties described by (\ref{eq:esym6d}), (\ref{eq:dell_N=4}), and (\ref{eq:6dlim_intro}), we conclude that the structure of the $q$-expansions is uniform for all the coefficients $\widehat{C}_{i_1,\dots,i_n}\brc{\epsilon,\tau,\hat\tau}$: each power of $q$ is multiplied by the Riemann theta functions to the power of $2N$ as in (\ref{eq:6dlim_intro}) and by the finite linear combination of non-positive powers of $\sigma\brc{\epsilon|\,\hat\tau}$ with coefficients being quasimodular forms in $\hat\tau$ with some particular weights. Thus, the exact expression for any given order in $q$ of any given function $\widehat{C}_{i_1,\dots,i_n}\brc{\epsilon,\tau,\hat\tau}$ can be computed.

\section{Double-elliptic Seiberg-Witten prepotential}
\label{sec:Dell_Prep}
According to \cite{AMM'16}, there exist non-linear equations for the Seiberg-Witten prepotential, which have exactly the $N$-particle double-elliptic system as its generic solution. With the help of these equations, the expression for the $N=3$ double-elliptic Seiberg-Witten prepotential was derived. After some minor simplifications, the obtained result can be generalized to the case of $N\geq 2$ as
\eqlb{eq:prepDellN}{\begin{array}{c}\ds
\mathcal{F}^{\textrm{Dell}}=\frac{\tau}2\sum_{i=1}^{N}a_i^2+
\frac{\epsilon^2}{2\pi\imath\,\beta^2}
\sum_{i<j}\log\,\theta_{11}\brc{\pi^{-1}\beta\,a_{ij}|\,\hat\tau}-\\
\ds
-\frac{\epsilon^2}{\pi\imath\,\beta^2}\sum_{\at{i_1,\dots,i_n\in\mathbb{Z}_{\geq0}}
{i_1+\dots+i_n\neq0}}\,\epsilon^{2\,i_1+\dots+2\,i_n}\,
\widehat{C}_{i_1,\dots,i_n}\brc{\epsilon,\tau,\hat\tau}\,
\prod^{n}_{\at{k=1}{\brc{\vec\alpha_k\in\Delta_+}}}\frac{1}
{\sigma\brc{\beta\,\vec\alpha_k\cdot\vec a\,|\,\hat\tau}^{2\,i_k}},
\end{array}}
where $n=N\brc{N-1}/2$, and $\Delta_+$ is the set of all positive roots $\bfi{\vec e_i-\vec e_j;\,i<j}$ in the $A_{N-1}$ root system.
The coefficients $\widehat{C}_{i_1,\dots,i_n}$ are fully symmetric under the permutation of indices $i_1,\dots,i_n$ and depend on the both elliptic parameters only through the Eisenstein series.
For example, $\widehat{C}_{i_1,\dots,i_n}$ can be decomposed in powers of $\epsilon$ in the following way:
\eqlb{eq:C_Dec_N}{\widehat{C}_{i_1,\dots,i_n}\brc{\epsilon,\tau,\hat\tau}
=\sum_{k=0}^{+\infty}\,\epsilon^{2k} \sum_{\at{m_1,m_2,m_3\geq 0}{m_1+2m_2+3m_3=k}}
\hat{E}_2^{m_1}\, \hat{E}_4^{m_2}\, \hat{E}_6^{m_3}\,
\widehat{C}_{i_1,\dots,i_n,k,\brc{m}}\brc{\tau},}
where $\brc{m}$ stands for the multi-index $(m_1,m_2,m_3)$,
\eq{\hat{E}_2\equiv E_2\brc{\hat\tau},\qq \hat{E}_4\equiv E_4\brc{\hat\tau}, \qq
\hat{E}_6\equiv E_6\brc{\hat\tau},}
and $\widehat{C}_{i_1,\dots,i_n,k,\brc{m}}\brc{\tau}$ are quasimodular forms of weight $2\,i_1+\dots+2\,i_n+2k$. Also, one should impose some additional restrictions on the summation over the indices $i_1,\dots,i_n$ in (\ref{eq:prepDellN}), since otherwise not all the coefficients $\widehat{C}_{i_1,\dots,i_n}\brc{\epsilon,\tau,\hat\tau}$ are independent: there are some relations between the functions $\sigma\brc{\beta\,\vec\alpha_k\cdot\vec a\,|\,\hat\tau}$.

Since the functions $\widehat{C}_{i_1,\dots,i_n,k,\brc{m}}\brc{\tau}$ are quasimodular forms, they can be realized as polynomials in the Eisenstein series $E_2$, $E_4$, and $E_6$:
\eq{E_2\brc{\tau}=1-24\sum _{n\in\mathbb{N}}\frac{n\,q^n}{1-q^n},}
\eq{E_4\brc{\tau}=1+240\sum _{n\in\mathbb{N}}\frac{n^3\,q^n}{1-q^n},}
\eq{E_6\brc{\tau}=1-504\sum _{n\in\mathbb{N}}\frac{n^5\,q^n}{1-q^n}.}
The constant terms in the expansions of $\widehat{C}_{i_1,\dots,i_n,k,\brc{m}}\brc{\tau}$ in powers of $q=\exp\brc{2\pi\imath\,\tau}$ correspond to the perturbative part of the prepotential $\mathcal{F}^{\textrm{Dell}}$.
The exact answer for the perturbative part is known and can be written in terms of the second derivatives as follows:
\eqlb{eq:pert_N}{
\def\arraystretch{3}
\begin{array}{c}
\ds i\neq j: \frac{\partial^2 \mathcal{F}^{\textrm{Dell}}}{\partial a_i \partial a_j}=
-\frac1{\pi\imath}\log\frac{\theta_{11}\brc{\pi^{-1}\epsilon|\,\hat\tau}}
{\pi^{-1}\,\theta_{11}'\brc{0|\,\hat\tau}}-
\frac1{2\pi\imath}\log\brc{\frac1{\sigma\brc{\epsilon|\,\hat\tau}^2}- \frac{1} {\sigma\brc{\beta\,a_{ij}|\,\hat\tau}^{2}}}+
\sum_{k\in\mathbb{N}}q^k\frac{\partial^2\mathcal{F}^{\brc{k}}}{\partial a_i \partial a_j},
\\ \ds
\frac{\partial^2 \mathcal{F}^{\textrm{Dell}}}{\partial a_i^2}=
\tau+\frac{N-1}{\pi\imath}\log\frac{\theta_{11}\brc{\pi^{-1}\epsilon|\,\hat\tau}}
{\pi^{-1}\,\theta_{11}'\brc{0|\,\hat\tau}}+
\frac1{2\pi\imath}\sum_{j\neq i}\log\brc{\frac1{\sigma\brc{\epsilon|\,\hat\tau}^2}- \frac{1} {\sigma\brc{\beta\,a_{ij}|\,\hat\tau}^{2}}}+
\sum_{k\in\mathbb{N}}q^k\frac{\partial^2\mathcal{F}^{\brc{k}}}{\partial a_i^2},
\end{array}}
where $a_{ij}\equiv a_i-a_j$ and the functions $\mathcal{F}^{\brc{k}}=\mathcal{F}^{\brc{k}}\brc{\textbf{a},\,\epsilon,\,\beta,\,\hat\tau}$ describing the instanton corrections do not depend on the first elliptic parameter $\tau$. As one can note, at the right-hand sides of (\ref{eq:pert_N}) there are some specific $a$-independent terms that are essential for the computation of the limit  $\sigma\brc{\epsilon|\,\hat\tau}\rightarrow\infty$.

\section{First modular anomaly equation}
\label{sec:First_MA}
From the M theory point of view, the curve $\Gamma^{\textrm{Dell}}$ is a two-dimensional Riemann surface in a compact four-dimensional manifold $X_{\brc{\epsilon,\hat\tau}}$ defined earlier in the introduction. $X_{\brc{\epsilon,\hat\tau}}$ can be thought of as a $\mathbb{T}^2$ bundle over $\Sigma$, where $\Sigma$ and $\mathbb{T}^2$ are two different tori with moduli $\tau$ and $\hat\tau$. Under the projection $X_{\brc{\epsilon,\hat\tau}}\rightarrow\Sigma$, the curve $\Gamma^{\textrm{Dell}}$ maps to $\Sigma$, and this gives rise to the interpretation of  $\Gamma^{\textrm{Dell}}$ as an $N$-sheeted covering of the base torus $\Sigma$. To get a proper geometrical description of this covering, one needs to determine the corresponding multivalued function from $\Sigma$ to $\mathbb{T}^2$. In the $4d$ case, when $\mathbb{T}^2$ is decompactified to a complex plane $\mathbb{C}$, the homology basis $\brc{A_i,B_i}$ for the curve $\Gamma^{\textrm{CM}}$ is given by the lifts $A_i, B_i$ of the cycles $A, B$ on the base $\Sigma$ to each sheet:
\eqlb{fig:CM_curve}{\includegraphics[height=7cm]{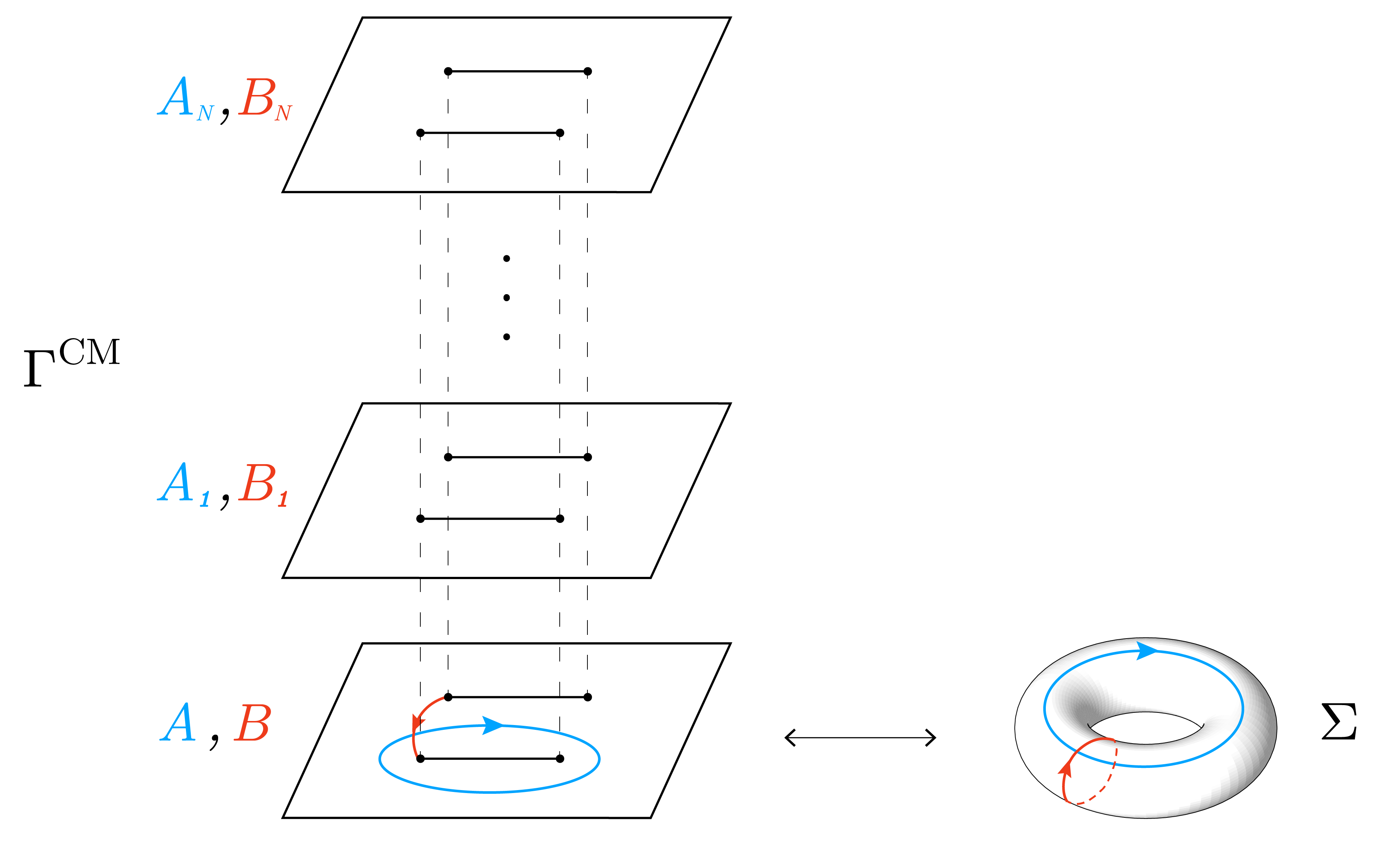}}
To draw a similar picture for the curve $\Gamma^{\textrm{Dell}}$, one needs to compactify each copy of $\mathbb{C}$ to a torus, which can be done, for example, by adding two cuts on each sheet.
However, the placement of the resulting four cuts is crucial and affects the basic properties of the curve, since some of the cuts might be coincident. Thus, instead of guessing the right geometrical interpretation, we use the explicit expression for the double-elliptic prepotential (\ref{eq:prepDellN}) and define the $N\times N$ period matrix of $\Gamma^{\textrm{Dell}}$ by
\eqlb{eq:6d_matrix}{T_{ij}=\frac{\partial^2 \mathcal{F}^{\textrm{Dell}}}{\partial a_i \partial a_j}.}
This implies that the homology basis for $\Gamma^{\textrm{Dell}}$ is still given by $\brc{A_i,B_i}$ and properties of the curve are described by the picture (\ref{fig:CM_curve}). In particular, (\ref{fig:CM_curve}) is very useful for understanding the properties of $T_{ij}$ with respect to the modular transformations of the first elliptic parameter $\tau$. On the other hand, the picture (\ref{fig:CM_curve}) is not applicable to description of the behavior of the second elliptic parameter $\hat\tau$ under the modular transformations of $\tau$. The same is true for the properties of $T_{ij}$ with respect to the modular transformations of $\hat\tau$. At this point, the explicit expression (\ref{eq:prepDellN}) comes into play.

Let us start with the modular transformations of the first elliptic parameter
\eqlb{tMod2}{\tau\rightarrow\tau+1\quad \textrm{and}\quad \tau\rightarrow-\frac1{\tau}.}
The first transformation is trivial and results in the following shift of the period matrix:
\eqlb{eq:t_t+1}{T_{ij}\xrightarrow{\tau\rightarrow\tau+1} T_{ij}+\delta_{ij}.}
The second transformation from (\ref{tMod2}) interchanges the cycles $A$ and $B$ on the base torus:
\eqlb{eq:Change_AB}{A \xrightarrow{\tau\rightarrow-1/\tau} B,\quad
B \xrightarrow{\tau\rightarrow-1/\tau}- A.}
Since the cycles $A_i$ and $B_i$ on each sheet of the covering are situated exactly above the cycles $A$ and $B$, (\ref{eq:Change_AB}) results in
\eq{A_i \xrightarrow{\tau\rightarrow-1/\tau} B_i,\quad
B_i \xrightarrow{\tau\rightarrow-1/\tau}- A_i.}
Taking into the account the definitions of the flat moduli $a_i$ and their duals $a^D_i$
\eqlb{eq:a_Def}{a_i=\frac1{2\pi\imath}\oint_{A_i} k\rmd z,\quad
a^D_i=\frac1{2\pi\imath}\oint_{B_i} k\rmd z=\frac{\partial \mathcal{F}^{\textrm{Dell}}}
{\partial a_i},}
one gets
\eqlb{eq:Mod_aT}{a_i \xrightarrow{\tau\rightarrow-1/\tau} a_i^D,\quad
a_i^D \xrightarrow{\tau\rightarrow-1/\tau}- a_i,\quad
T \xrightarrow{\tau\rightarrow-1/\tau} -T^{-1}.}
The transformations for the other parameters can be written in terms of yet unknown functions $\epsilon'$, $\beta'$, $\hat\tau'$:
\eq{\epsilon\xrightarrow{\tau\rightarrow-1/\tau} \epsilon',\quad
\beta\xrightarrow{\tau\rightarrow-1/\tau} \beta',\quad
\hat\tau\xrightarrow{\tau\rightarrow-1/\tau}\hat\tau'.}
In what follows, we treat $\epsilon'$, $\beta'$, $\hat\tau'$ as series expansions in powers of $\epsilon$ and require, that the coefficients in these expansions do not depend on the flat moduli $\textbf{a}$.

The modular transformations (\ref{eq:Mod_aT}) and the quasimodular properties of the coefficients in the series expansion (\ref{eq:prepDellN}) allow one to determine $\epsilon'$, $\beta'$, $\hat\tau'$. To this end, we reformulate the transformation of the period matrix $T\brc{\textbf{a},\,\epsilon,\,\beta,\,\tau,\,\hat\tau}$ as
\eqlb{pM_eq_N}{T\brc{\textbf{a}^D,\,\epsilon',\,\beta',\,-\frac1{\tau},\,\hat\tau'} =-T\brc{\textbf{a},\,\epsilon,\,\beta,\,\tau,\,\hat\tau}^{-1}.}
The latter equation can be solved perturbatively in each order in $\epsilon$ for particular values of $N$. Evaluation for $N=2,3,4$ demonstrates that the solution is very simple and, in terms of $\epsilon'$, $\beta'$, and $\hat\tau'$, can be represented as:
\eqlb{eq:e'b't'}{\epsilon'=\frac{\epsilon}{\tau},\quad
\beta'=\frac{\beta}{\tau},\quad
\hat\tau'=\hat\tau -\frac{N\,\epsilon^2}{\pi^2\,\tau}.}
There are different ways to confirm that (\ref{eq:e'b't'}) is a proper general solution. A straightforward way is to solve (\ref{pM_eq_N}) for higher values of $N$. An easier way is to consider the first modular anomaly equation, which is introduced below in (\ref{eq:mod_amomaly_6d_N}), and solve it perturbatively in $\epsilon$. Then, in the first non-zero order, the appearance of $N$ in the function $\hat\tau'$ is necessary to ensure the consistency of the equation. Summarizing the results, we describe the action of the second modular transformation from (\ref{tMod2}) as
\eqlb{eq:t_mod2}{\boxed{\tau\rightarrow -\frac1{\tau},\quad
\epsilon\rightarrow \frac{\epsilon}{\tau},\quad
\beta\rightarrow \frac{\beta}{\tau},\quad
\hat\tau\rightarrow\hat\tau -\frac{N\,\epsilon^2}{\pi^2\,\tau},\quad
a_i\rightarrow a_i^{D},\quad
T\rightarrow - T^{-1}.}}
An interesting feature of (\ref{eq:t_mod2}) to pay attention is the transformation law of the parameter $\beta$. As we explained in the introduction, the $\beta$ inverse is proportional to the radius of the forth spacetime dimension: $\beta^{-1}=2 R_4$. Therefore, the natural transformation for $\beta$ under the modular transformation $\hat\tau\rightarrow-1/\hat\tau$ of the second elliptic parameter $\hat\tau=\imath\, R_5/ R_4$ is $\beta\rightarrow\beta/\hat\tau$ and this will be the case in the next section. The fact that we have $\beta\rightarrow\beta/\tau$ under the modular transformation $\tau\rightarrow-1/\tau$ of the first elliptic parameter could mean that
one of the cycles of $\Sigma$ is mapped onto one of the cycles of $\mathbb{T}^2$ and some of the four cuts on each sheet of $\Gamma^{\textrm{Dell}}$ coincide in accordance with our earlier assumptions.

To derive the $6d$ modular anomaly equation, consider the linear combination of the flat moduli
\eq{a_i^{D}-\tau\,a_i=\frac{\partial \wt{\mathcal{F}}}{\partial a_i},\quad
\wt{\mathcal{F}}\equiv\mathcal{F^{\textrm{Dell}}}-\frac{\tau}2\sum_{i=1}^{N}a_i^2.}
With the help of (\ref{eq:Mod_aT}), we obtain
\eq{a_i^{D}-\tau\,a_i\xrightarrow{\tau\rightarrow-1/\tau}
\frac1{\tau}\brc{a_i^{D}-\tau\,a_i}}
or
\eqlb{eq:wtF_Mod}{\frac{\partial \wt{\mathcal{F}}}{\partial a_i} \brc{\textbf{a}^D,\,\frac{\epsilon}{\tau},\,\frac{\beta}{\tau},\,-\frac1{\tau}, \,\hat\tau -\frac{N\,\epsilon^2}{\pi^2\,\tau}}= \frac1{\tau}
\frac{\partial \wt{\mathcal{F}}}{\partial a_i} \brc{\textbf{a},\,\epsilon,\, \beta,\, \tau, \,\hat\tau}.}
Function $\wt{\mathcal{F}}$ is the sum of the perturbative and instanton parts of the prepotential and depends on the elliptic parameter $\tau$ only through the Eisenstein series $E_2$, $E_4$, $E_6$ with the following modular properties:
\eq{E_2\brc{-\frac1{\tau}}=\tau^2 E_2\brc{\tau}+\frac{6\,\tau}{\pi\imath},\quad
E_4\brc{-\frac1{\tau}}=\tau^4 E_4\brc{\tau},\quad
E_6\brc{-\frac1{\tau}}=\tau^6 E_6\brc{\tau}.}
As it can be seen from  (\ref{eq:prepDellN}) and (\ref{eq:C_Dec_N}), $\wt{\mathcal{F}}$ possesses some type of scaling invariance
\eq{\wt{\mathcal{F}} \brc{\textbf{a}^D,\,\frac{\epsilon}{\tau},\,\frac{\beta}{\tau},\,\tau^2 E_2+\frac{6\,\tau}{\pi\imath},\,\tau^4 E_4,\,\tau^6 E_6,\,\hat\tau -\frac{N\,\epsilon^2}{\pi^2\,\tau}}=\wt{\mathcal{F}} \brc{\frac{\textbf{a}^D}{\tau},\,\epsilon,\,\beta,\,E_2+\frac{6}{\pi\imath\,\tau},\, E_4,\, E_6,\,\hat\tau-\frac{N\,\epsilon^2}{\pi^2\,\tau}}.}
This allows us to simplify (\ref{eq:wtF_Mod}):
\eqlb{eq:wF_symm}{\frac{\partial \wt{\mathcal{F}}}{\partial a_i} \brc{\textbf{a}+\frac1{\tau}\nabla_{\textbf{a}}\,\wt{\mathcal{F}},\,E_2+\frac{6}{\pi\imath\,\tau}, \,\hat\tau-\frac{N\,\epsilon^2}{\pi^2\,\tau}}= \frac{\partial \wt{\mathcal{F}}}{\partial a_i}
\brc{\textbf{a},\,E_2,\,\hat\tau},}
where $\nabla_{\textbf{a}}=\brc{\partial/ \partial a_1,\dots,\partial/ \partial a_N}$ and the dependence on the other arguments is implied on the both sides of the equality. This equation manifest the new symmetry of the function $\wt{\mathcal{F}}$ and describes the dependence of the prepotential on the second Eisenstein  series $E_2\brc{\tau}$. Consider the first order in the expansion of (\ref{eq:wF_symm}) in powers of $1/\tau$:
\eq{\sum_{j=1}^{N}\frac{\partial \wt{\mathcal{F}}}{\partial a_j}
\frac{\partial^2 \wt{\mathcal{F}}}{\partial a_i\, \partial a_j} +\frac{6}{\pi\imath}
\frac{\partial^2 \wt{\mathcal{F}}}{\partial a_i\,\partial E_2} -\frac{N\,\epsilon^2}{\pi^2}
\frac{\partial^2 \wt{\mathcal{F}}}{\partial a_i\,\partial \hat\tau}=0.}
Integrating with respect to $a_i$ and omitting the constant of integration, we obtain the $6d$ generalization of the MNW \cite{MNW} modular anomaly equation:
\eqlb{eq:mod_amomaly_6d_N}{\boxed{
\frac{6}{\pi\imath}\frac{\partial \mathcal{F}^{\textrm{Dell}}}{\partial E_2} - \frac{N\epsilon^2}{\pi^2} \frac{\partial \mathcal{F}^{\textrm{Dell}}}{\partial \hat\tau}=
-\frac{1}{2}\sum_{i=1}^{N}\brc{\frac{\partial \mathcal{F}^{\textrm{Dell}}}{\partial a_i}-\tau\,a_i}^2.}}

\section{Second modular anomaly equation}
\label{sec:Second_MA}
We learned in the previous section that the two tori $\Sigma$ and $\mathbb{T}^2$ play different roles in the geometrical description of the curve $\Gamma^{\textrm{Dell}}$. In particular, the definitions of the moduli $\textbf{a}$ and $\textbf{a}^{D}$ are essentially connected with the cycles $A$ and $B$ on the base torus $\Sigma$, and the period matrix $T$ has the $U(1)$-decoupling property:
\eq{\sum_{i=1}^{N}T_{ij}=\tau,\quad \forall j=1,\dots,N.}
This indicates that the theory should behave differently under the modular transformations of the first and the second elliptic parameters.

In order to understand the behavior of the period matrix $T$ under the modular transformations of the second elliptic parameter $\hat\tau$, we first consider it at the classical and perturbative levels. With the help of the exact expressions (\ref{eq:pert_N}) and of the expansion
\eqlb{eq:logthetaDec2}{\log\frac{\theta_{11}\brc{\pi^{-1}z|\,\hat\tau}}
{\pi^{-1}z\,\theta_{11}'\brc{0|\,\hat\tau}}=-\sum_{k=1}^{+\infty} \frac{\zeta\brc{2k}}{k\,\pi^{2k}}\,E_{2k}\brc{\hat\tau}\,z^{2k},}
we establish that the sum of the classical and perturbative parts of the period matrix is invariant under the following transformations of the moduli:
\eqlb{eq:moduli2_trans1}{\boxed{
\hat\tau\rightarrow \hat\tau+1,\quad
\epsilon\rightarrow \epsilon,\quad
\beta\rightarrow \beta,\quad
\tau\rightarrow\tau,\quad
a_i\rightarrow a_i,\quad T_{ij}\rightarrow T_{ij}} }
and
\eqlb{eq:moduli2_trans2}{\boxed{
\hat\tau\rightarrow -\frac1{\hat\tau},\quad
\epsilon\rightarrow \frac{\epsilon}{\hat\tau},\quad
\beta\rightarrow \frac{\beta}{\hat\tau},\quad
\tau\rightarrow\tau -\frac{N\,\epsilon^2}{\pi^2\,\hat\tau},\quad
a_i\rightarrow a_i,\quad T_{ij}\rightarrow T_{ij}-\frac{\epsilon^2}{\pi^2\,\hat\tau}.}}
In fact, the second modular transformation (\ref{eq:moduli2_trans2}) shifts the period matrix. However, this shift can be removed by adding to the classical part of the prepotential a term proportional to $\brc{\sum_i a_i}^2$. This term is also relevant for the computation of the limit $\sigma\brc{\epsilon|\,\hat\tau}\rightarrow\infty$, which we will discuss in the next section.

We notice that the transformations (\ref{eq:moduli2_trans1}) and (\ref{eq:moduli2_trans2}) do not mix the instanton part of the period matrix with the classical and perturbative parts. Thus, it is natural to assume that the instanton part is also invariant under the modular transformations of the second elliptic parameter (\ref{eq:moduli2_trans1}) and (\ref{eq:moduli2_trans2}). This assumption provides us with a non-trivial equation on the prepotential, which can be reformulated in terms of the linear relations between the functions $\widehat{C}_{i_1,\dots,i_n}\brc{\epsilon,\tau,\hat\tau}$ and their derivatives. We derive later the exact expressions for some first functions $\widehat{C}_{i_1,\dots,i_n,k,\brc{m}}\brc{\tau}$ in the cases when $N=2,3$ and the relations will be valid for all the computed expressions. We consider this as a strong evidence in favor of the assumption being made. A less direct evidence is provided by the fact that the transformation laws for the parameters $\epsilon$, $\beta$, $\tau$, and $\hat\tau$ are covariant under the permutation of $\tau$ and $\hat\tau$.

The invariance of the period matrix under the modular transformations of the second elliptic parameter imply that the following equations on the period matrix hold:
\eqlb{eq:TMod2_N}{
\def\arraystretch{2}
\begin{array}{c}
\ds T_{ij}\brc{\textbf{a},\,\epsilon,\,\beta, \,\tau,\,\hat\tau+1} = T_{ij}\brc{\textbf{a},\,\epsilon,\,\beta,\,\tau,\,\hat\tau},\\
\ds T_{ij}\brc{\textbf{a},\,\frac{\epsilon}{\hat\tau},\, \frac{\beta}{\hat\tau}, \,\tau -\frac{N\,\epsilon^2}{\pi^2\,\hat\tau},\, -\frac1{\hat\tau}} = T_{ij}\brc{\textbf{a},\,\epsilon,\,\beta,\,\tau,\,\hat\tau}-\frac{\epsilon^2}{\pi^2\,\hat\tau}.
\end{array}}
Since the period matrix depends on the second elliptic parameter only through the Eisenstein series, the first equation from (\ref{eq:TMod2_N}) is trivial.
The second equation from (\ref{eq:TMod2_N}) gives
\eqlb{eq:dFSym2}{\frac{\partial}{\partial a_i} \brc{ \mathcal{F}^{\textrm{Dell}}\brc{\textbf{a},\,\frac{\epsilon}{\hat\tau},\, \frac{\beta}{\hat\tau}, \,\tau -\frac{N\,\epsilon^2}{\pi^2\,\hat\tau},\, -\frac1{\hat\tau}} - \mathcal{F}^{\textrm{Dell}}\brc{\textbf{a},\,\epsilon,\,\beta,\,\tau,\,\hat\tau}}=
-\frac{\epsilon^2}{\pi^2\,\hat\tau}\sum_{j=1}^{N}a_j.}
Taking into the account the scaling properties with respect to the second elliptic parameter
\eq{\mathcal{F}^{\textrm{Dell}}\brc{\textbf{a},\,\frac{\epsilon}{\hat\tau},\, \frac{\beta}{\hat\tau}, \,\tau -\frac{N\,\epsilon^2}{\pi^2\,\hat\tau},\,\hat\tau^2 \hat E_2+ \frac{6\,\hat\tau}{\pi\imath}, \,\hat\tau^4 \hat E_4,\,\hat\tau^6 \hat E_6}=\mathcal{F}^{\textrm{Dell}}\brc{\textbf{a}, \,\epsilon,\,\beta,\, \tau -\frac{N\,\epsilon^2}{\pi^2\,\hat\tau}, \,\hat E_2 +\frac{6}{\pi\imath\,\hat\tau},\, \hat E_4,\, \hat E_6},}
we rewrite (\ref{eq:dFSym2}) as
\eqlb{eq:dFSym_E2}{\frac{\partial}{\partial a_i} \brc{ \mathcal{F}^{\textrm{Dell}}\brc{\textbf{a},\,\epsilon,\,\beta, \,\tau -\frac{N\,\epsilon^2}{\pi^2\,\hat\tau},\, \hat E_2 +\frac{6}{\pi\imath\,\hat\tau}} - \mathcal{F}^{\textrm{Dell}}\brc{\textbf{a},\,\epsilon,\,\beta,\,\tau,\,\hat E_2}}=
-\frac{\epsilon^2}{\pi^2\,\hat\tau}\sum_{j=1}^{N}a_j.}
This leads us to the second modular anomaly equation:
\eqlb{eq:mod_amomaly_N2}{\boxed{
\frac{6}{\pi\imath}\frac{\partial \mathcal{F}^{\textrm{Dell}}}{\partial \hat E_2} - \frac{N\,\epsilon^2}{\pi^2} \frac{\partial \mathcal{F}^{\textrm{Dell}}}{\partial \tau}=
-\frac{\epsilon^2}{2\,\pi^2}\brc{\sum_{i=1}^{N}a_i}^2,}}
where the integration constant coming from (\ref{eq:dFSym_E2}) was omitted.

Expanding (\ref{eq:mod_amomaly_N2}) in powers of $\epsilon$, in the first nonzero order we get the equation
\eq{-\frac{3\,\epsilon^2}{2\pi^2\beta^2} \frac{\partial}{\partial \hat E_2} \sum_{i<j}\log\theta_{11}\brc{\pi^{-1}\beta\,a_{ij}|\,\hat{\tau}}
-\frac{N\,\epsilon^2}{2\,\pi^2} \sum_{i=1}^{N}a_i^2= -\frac{\epsilon^2}{2\,\pi^2}\brc{\sum_{i=1}^{N}a_i}^2,}
which can be checked with the help of (\ref{eq:logthetaDec2}), and the particular value of the Riemann zeta function $\zeta\brc{2}=\pi^2/6$. As it was mentioned earlier, the second modular anomaly equation (\ref{eq:mod_amomaly_N2}) can be solved in each order in $\epsilon$, and the solution is given by the linear relations between the functions $\widehat{C}_{i_1,\dots,i_n}\brc{\epsilon,\tau,\hat\tau}$ and their derivatives. In the simplest case of $N=2$, the exact solution is given by the recurrence relations (\ref{eq:mod_anomaly2_Rec1}).

\section{The limit $\sigma\brc{\epsilon|\,\hat\tau}\rightarrow\infty$}
\label{sec:mass_lim}
Before we start solving the first modular anomaly equation (\ref{eq:mod_amomaly_6d_N}), it is useful to understand the limit of $\sigma\brc{\epsilon|\,\hat\tau}\rightarrow\infty$.
In the $4d$ case, the limit of infinite mass $m$ (or the pure gauge limit) was very well studied \cite{HPh'98,MNW}. At the same time, as $m$ goes to infinity, one should simultaneously bring the elliptic parameter $\tau$ to imaginary infinity, so that the resulting cutoff $\Lambda$ is finite:
\eq{m\rightarrow\infty,\quad \textrm{Im}\,\tau \rightarrow+\infty,\quad
m^{2N}\exp\brc{2\pi\imath\,\tau}\rightarrow\brc{-1}^N\Lambda^{2N}.}
The limit of the $4d$ prepotential can be described as an infinite series in powers of $\Lambda$ and corresponds to the periodic Toda integrable system \cite{GKMMM'95,MW'96}. Analogous considerations in the $6d$ case could result in the elliptic generalization of the Toda system \cite{AS'97,Kri'99}.

The pure gauge limit is replaced in the $6d$ theory by the special point $\epsilon=\epsilon^{\infty}$, which is defined by (\ref{eq:einf_intro}). The fact that there is no limit of infinite mass in the $6d$ case, is clear from both the M theory and the pure analytical point of view. The M theory viewpoint was described in the introduction, and it relies on the interpretation of the mass parameter as a shift in two compact dimensions. In the pure analytical approach, one would consider the exact expression (\ref{eq:pert_N}) for the perturbative part of the period matrix and notice that it depends on the elliptic function of $\epsilon$, which does not have a limit, when $\epsilon$ tends to a real or imaginary infinity. Since $\epsilon^{\infty}$ is just some finite value of $\epsilon$, one might argue that there is no need to approach this value in the special limit of $\epsilon\rightarrow\epsilon^{\infty}$. However, the elliptic function $\sigma\brc{\epsilon|\,\hat\tau}$ has a pole at $\epsilon=\epsilon^{\infty}$ and it is a non-trivial requirement for the prepotential to have a well-defined limit at $\epsilon\rightarrow\epsilon^{\infty}$. We treat this limit by analogy with the pure gage limits of the $4d$ and $5d$ theories. The complexified coupling constant $\tau$ and the mass parameter $\epsilon$ should be replaced by one new parameter $\hat\Lambda$. To this end, the following shift of the period matrix is required:
\eq{\forall i,j:\quad
T^{\infty}_{ij}=T_{ij}-\frac{\tau}{N}.}
This shift corresponds to an additional classical term in the prepotential:
\eq{\mathcal{F}^{\infty}=\mathcal{F}^{\textrm{Dell}} - \frac{\tau}{2N}\brc{\sum_{i=1}^{N}a_i}^2.}
Then, a counterpart of the pure gauge limit in the $6d$ theory is
\eqlb{eq:6d_limit}{\boxed{
\epsilon\rightarrow\epsilon^{\infty},\quad
\sigma\brc{\epsilon|\,\hat\tau}\rightarrow\infty, \quad
\frac{\theta_{11}\brc{\pi^{-1}\epsilon |\,\hat\tau}^{2N}}
{\pi^{-2N}\, \theta_{11}'\brc{0|\,\hat\tau}^{2N}} \exp\brc{2\pi\imath\,\tau}\rightarrow \brc{-1}^N\hat\Lambda^{2N}.}}
For the period matrix one gets
\eqlb{eq:dN_lim}{
\def\arraystretch{3}
\begin{array}{c}
\ds \lim_{\epsilon\rightarrow\epsilon^{\infty}} T^{\infty}_{ii} =
\frac1{2\pi\imath}\sum_{j\neq i} \log\brc{ \frac{\hat\Lambda^2} {\sigma\brc{\beta\,a_{ij}|\,\hat\tau}^{2}}} +
\lim_{\epsilon\rightarrow\epsilon^{\infty}}
\sum_{k\in\mathbb{N}}q^k\frac{\partial^2\mathcal{F}^{\brc{k}}}{\partial a_i^2},
\\
\ds i\neq j: \lim_{\epsilon\rightarrow\epsilon^{\infty}} T^{\infty}_{ij}= -
\frac1{2\pi\imath}\log\brc{\frac{\hat\Lambda^2} {\sigma\brc{\beta\,a_{ij}|\,\hat\tau}^{2}}}+
\lim_{\epsilon\rightarrow\epsilon^{\infty}} \sum_{k\in\mathbb{N}}q^k\frac{\partial^2\mathcal{F}^{\brc{k}}}{\partial a_i \partial a_j}.
\end{array}}
As in the $4d$ case, the convergency condition for the instanton part of (\ref{eq:dN_lim}) imposes additional restrictions on the coefficients in the series expansion of the prepotential (\ref{eq:prepDellN}).
In order to satisfy this condition, we rewrite the instanton part as a power series in the new parameters $\nu\equiv\sigma\brc{\epsilon|\,\hat\tau}$ and $\wt q\equiv \exp\brc{2\pi\imath\,\wt\tau}$, where $\wt\tau$ is
\eqlb{eq:wtau}{\wt\tau=\tau+\frac{N}{\pi\imath}\log\frac{\pi\,\theta_{11}\brc{\pi^{-1}\epsilon|\,\hat\tau}}
{\sigma\brc{\epsilon|\,\hat\tau}\theta_{11}'\brc{0|\,\hat\tau}}.}
Then, the coefficients in the expansion (\ref{eq:prepDellN}) transform as
\eqlb{eq:lim_trans}{\def\arraystretch{3}
\begin{array}{c}
\ds \sum_{k=0}^{+\infty} \sum_{\at{m_1,m_2,m_3\geq 0}{m_1+2m_2+3m_3=k}}
\epsilon^{2\brc{i_1+\dots+i_n+k+1}}
\hat{E}_2^{m_1}\, \hat{E}_4^{m_2}\, \hat{E}_6^{m_3}\, \widehat{C}_{i_1,\dots,i_n,k,\brc{m}}\brc{\tau} =
\\ \ds
=\sum_{k=0}^{+\infty} \sum_{\at{m_1,m_2,m_3\geq 0}{m_1+2m_2+3m_3=k}}
\nu^{2\brc{i_1+\dots+i_n+k+1}}
\hat{E}_2^{m_1}\, \hat{E}_4^{m_2}\, \hat{E}_6^{m_3}\, \widehat{C}^{\infty}_{i_1,\dots,i_n,k,\brc{m}}\brc{\wt\tau}
\end{array}}
and the functions $\widehat{C}^{\infty}_{i_1,\dots,i_n,k,\brc{m}}$ are linear combinations of $\widehat{C}_{i_1,\dots,i_n,k,\brc{m}}$ and their derivatives. In terms of these new parameters, the limit (\ref{eq:6d_limit}) can be described as $\nu^{2N}\wt q\rightarrow\brc{-1}^N\hat\Lambda^{2N}$. Since there should be no divergent terms $\nu^{2Nm+k}\,\wt q^{\,m}$ with $m,\,k\in\mathbb{N}$ at the {\it r.h.s.} of (\ref{eq:lim_trans}), we get the following restrictions on the functions $\widehat{C}^{\infty}_{i_1,\dots,i_n,k,\brc{m}}$:
\eqlb{eq:wtc_cons}{\widehat{C}^{\infty}_{i_1,\dots,i_n,k,\brc{m}}\brc{\wt\tau} = \sum_{l=0}^{+\infty}\widehat{C}^{\infty}_{i_1,\dots,i_n,k,\brc{m},l}\,\wt q^{\,l} = \widehat{C}^{\infty}_{i_1,\dots,i_n,k,\brc{m},0} + \sum_{l\geq\brc{i_1+\dots+i_n+k+1}/N} \widehat{C}^{\infty}_{i_1,\dots,i_n,k,\brc{m},l}\,\wt q^{\,l},}
where the first $\lceil \frac{i_1+\dots+i_n+k+1}2 \rceil-1$ coefficients $\widehat{C}^{\infty}_{i_1,\dots,i_n,k,\brc{m},l}$, $l\in \mathbb{N}$ should vanish, and the constant term $\widehat{C}^{\infty}_{i_1,\dots,i_n,k,\brc{m},0}$ was already taken into account in the limit (\ref{eq:dN_lim}) of the perturbative part of the prepotential.

\section{Modular anomaly for $N=2$}
\label{sec:N=2}
In this section, we consider the two-particle double-elliptic integrable system in the center of mass frame ($a_1+a_2=0$). This case is the simplest one from the computational point of view, and, at the same time, it reflects all the relevant phenomena arising in the general $N$-particle case.
The corresponding prepotential can be written as
\eqlb{eq:prep_N=2}{\begin{array}{c}\ds
\mathcal{F}^{\textrm{Dell}}=\tau\frac{a^2}2+
\frac{\epsilon^2}{4\pi\imath\,\beta^2}
\log\theta_{11}\brc{2\,\pi^{-1}a\,\beta|\,\hat{\tau}}-\\
\ds
-\frac{\epsilon^2}{8\pi\imath\,\beta^2}
\sum_{n=1}^{+\infty}\epsilon^{2n}\,\hat c_{n} \brc{\epsilon, \tau, \hat\tau}
\frac{1}{\sigma\brc{2\,a\,\beta|\,\hat\tau}^{2n}},
\end{array}
}
where
\eqlb{eq:cn_Dec}{\hat c_{n} \brc{\epsilon,\tau,\hat\tau}=
\sum_{k=0}^{+\infty}\,\epsilon^{2k} \sum_{\at{m_1,m_2,m_3\geq 0}{m_1+2m_2+3m_3=k}}
\hat{E}_2^{m_1}\, \hat{E}_4^{m_2}\, \hat{E}_6^{m_3}\, \hat c_{nk\brc{m}}\brc{\tau}.}
The second derivative of the prepotential (\ref{eq:prep_N=2}) defines the period matrix:
\eqlb{eq:pert_N=2}{T=\frac{\partial^2 \mathcal{F}^{\textrm{Dell}}}{\partial a^2}=
\tau+\frac2{\pi\imath}\log\frac{\theta_{11}\brc{\pi^{-1}\epsilon|\,\hat\tau}}
{\pi^{-1}\,\theta_{11}'\brc{0|\,\hat\tau}}+
\frac1{\pi\imath}\log\brc{\frac1{\sigma\brc{\epsilon|\,\hat\tau}^2}- \frac{1} {\sigma\brc{2\,a\,\beta|\,\hat\tau}^{2}}}+
\sum_{k\in\mathbb{N}}q^k\frac{\partial^2\mathcal{F}^{\brc{k}}}{\partial a^2}.}
The modular transformations of the first elliptic parameter $\tau$ act on the period matrix as
\eqlb{eq:Modt_N=2}{
\def\arraystretch{2}
\begin{array}{c}
\ds T\brc{a,\,\epsilon,\,\beta,\,\tau+1,\,\hat\tau}= T\brc{a,\,\epsilon,\,\beta,\,\tau,\,\hat\tau}+1,
\\ \ds
T\brc{a^D,\,\frac{\epsilon}{\tau},\,\frac{\beta}{\tau},\,-\frac1{\tau},\,\hat\tau -\frac{2\,\epsilon^2}{\pi^2\,\tau}} =-T\brc{a,\,\epsilon,\,\beta,\,\tau,\,\hat\tau}^{-1}.
\end{array}}
The first modular anomaly equation is
\eqlb{eq:mod_amomaly_6d_N=2}{N=2:\quad
\frac{6}{\pi\imath}\frac{\partial \mathcal{F}^{\textrm{Dell}}}{\partial E_2} - \frac{2\epsilon^2}{\pi^2} \frac{\partial \mathcal{F}^{\textrm{Dell}}}{\partial \hat\tau}=
-\frac{1}{2}\brc{\frac{\partial \mathcal{F}^{\textrm{Dell}}}{\partial a}-\tau\,a}^2.}
It can be reformulated in terms of the recurrence relations for the functions $\hat c_{n}\brc{\epsilon,\tau,\hat\tau}$. Restoring the integration constant
in (\ref{eq:mod_amomaly_6d_N=2}) and using the standard differential equations for the function $\sigma\brc{z|\,\hat\tau}$ (see \ref{sec:sigma}), we get the set of relations
\eqlb{eq:Rec_in_N=2}{3\frac{\partial \hat c_1}{\partial E_2}+\frac{\epsilon^2}{\pi\imath}
\frac{\partial \hat c_1}{\partial \hat\tau} -\frac12 + \epsilon^2 \hat E_2\,\hat c_1-
\frac{\epsilon^4}6 \brc{\hat E_2^2-\hat E_4}\brc{\hat c_1^2+2\,\hat c_2} +
\frac{2\,\epsilon^6}{27}\brc{\hat E_2^3-3\hat E_2 \hat E_4+2\hat E_6} \hat c_1 \hat c_2=0,}
\eqlb{eq:Rec_N=2}{\begin{array}{c}
\ds n\geq2:\\
\ds 3\frac{\partial \hat c_n}{\partial E_2}+ \frac{\epsilon^2}{\pi\imath}
\frac{\partial \hat c_n}{\partial \hat\tau}- \brc{n-1}\hat c_{n-1}
-\frac12 \sum_{m=1}^{n-2} m\brc{n-m-1}\hat c_{m}\, \hat c_{n-m-1}+ n\,\epsilon^2
\hat E_2\, \hat c_n - \frac{n+1}6\, \epsilon^4 \brc{\hat E_2^2-\hat E_4}\hat c_{n+1} +\\
\ds +\frac{\epsilon^2}2 \hat E_2 \sum_{m=1}^{n-1} m\brc{n-m} \hat c_{m}\, \hat c_{n-m} -\frac{\epsilon^4}6 \brc{\hat E_2^2-\hat E_4}
\sum_{m=1}^{n} m\brc{n-m+1} \hat c_{m}\, \hat c_{n-m+1}+\\
\ds + \frac{\epsilon^6}{54} \brc{\hat E_2^3-3\hat E_2 \hat E_4+2\hat E_6} \sum_{m=1}^{n+1} m\brc{n-m+2} \hat c_{m}\, \hat c_{n-m+2}=0.
\end{array}}
These relations are somewhat similar to the AMM/EO topological recursion \cite{AMM1'07,AMM2'07,AMM'09,EO'07,O'08} and can be solved exactly, if the proper boundary conditions are imposed. For example, one can use the convergency condition for the limit described in section \ref{sec:mass_lim}.

The second modular anomaly equation is
\eqlb{eq:mod_amomaly_2}{N=2:\quad
\frac{6}{\pi\imath}\frac{\partial \mathcal{F}^{\textrm{Dell}}}{\partial \hat E_2} - \frac{2\,\epsilon^2}{\pi^2} \frac{\partial \mathcal{F}^{\textrm{Dell}}}{\partial \tau}=0.}
Solving (\ref{eq:mod_amomaly_2}) for all orders in $\epsilon$, we get the following recurrence relations
\eqlb{eq:mod_anomaly2_Rec1}{n\geq1:\quad
3\,\frac{\partial \hat c_{n}}{\partial \hat E_2}+ \frac{\epsilon^2}{\pi\imath}
\frac{\partial \hat c_{n}}{\partial \tau} +\brc{n+1}\epsilon^2 \hat c_{n+1}=0}
or, equivalently,
\eqlb{eq:mod_anomaly2_Rec2}{n\geq1:\quad
\hat c_n=\frac{\brc{-1}^{n-1}}{n!}
\brc{\frac3{\epsilon^2}\frac{\partial}{\partial \hat E_2} +\frac1{\pi\imath} \frac{\partial}{\partial \tau}}^{n-1}\hat c_1.}
Thus, the second modular anomaly equation reduces the problem of evaluation of the $N=2$ double-elliptic prepotential to the problem of finding of one single function $\hat c_1 \brc{\epsilon, \tau, \hat\tau}$.

\subsection{Evaluating functions  $\hat c_{n}\brc{\epsilon,\tau,\hat\tau}$}
Equations (\ref{eq:Modt_N=2}) and (\ref{eq:mod_amomaly_6d_N=2}) along with the convergency condition for the limit of $\sigma\brc{\epsilon|\,\hat\tau}\rightarrow\infty$ can be used to define the instanton part of the $N=2$ prepotential completely, without making any additional assumptions about the functions $\hat c_{n}\brc{\epsilon,\tau,\hat\tau}$. The equations for the period matrix (\ref{eq:Modt_N=2}) describe the quasimodular properties of the coefficients $\hat c_{nk\brc{m}}\brc{\tau}$ in the expansion (\ref{eq:cn_Dec}). Then, the modular anomaly equation (\ref{eq:mod_amomaly_6d_N=2}) allows one to compute the dependence of $\hat c_{n}\brc{\epsilon,\tau,\hat\tau}$ on the second Eisenstein series $E_2\brc{\tau}$. Finally, the convergency condition from section \ref{sec:mass_lim} provides us with $E_2$-independent part of the $N=2$ prepotential. To simplify the notation, we replace the multi-index $\brc{m}$ in (\ref{eq:cn_Dec}) by an ordinary index $m$ for some first orders in $\epsilon$:
\eq{\hat c_{n} \brc{\epsilon,\tau,\hat\tau}=\hat c_{n00}+\epsilon^2 \hat E_2\,\hat c_{n10} +\epsilon^4\brc{\hat E_2^2\,\hat c_{n20}+\hat E_4\,\hat c_{n21}}+\epsilon^6\brc{\hat E_2^3\,\hat c_{n30}+\hat E_2\hat E_4\,\hat c_{n31}+ \hat E_6 \, \hat c_{n32}}+O\brc{\epsilon^8}.}

\subsubsection*{Quasimodular properties of $\hat c_{nk\brc{m}}\brc{\tau}$}
Expanding the second equation from (\ref{eq:Modt_N=2}) in powers of $\epsilon$, we obtain in the first nonzero order:
\eq{\hat c_{100}\brc{-\tau^{-1}}=\tau^2\,\hat c_{100}\brc{\tau}+\frac{\tau}{\pi\imath}.}
The first equation from (\ref{eq:Modt_N=2}) is equivalent to the periodicity condition $\hat c_{nkm}\brc{\tau+1}=\hat c_{nkm}\brc{\tau}$, which gives
\eqlb{eq:c100}{\hat c_{100}\brc{\tau}=\frac{E_2\brc{\tau}}6.}
In the next order, we get
\eqlb{eq:c110}{\hat c_{110}\brc{-\tau^{-1}}= \tau^4\,\hat c_{110}\brc{\tau} -\frac{2\tau^3}{\pi\imath}\hat c_{100}\brc{\tau} +\frac{\tau^2}{\pi^2},}
\eqlb{eq:c200}{\hat c_{200}\brc{-\tau^{-1}}= \tau^4\,\hat c_{200}\brc{\tau}+ \frac{2\tau^3}{\pi\imath}\hat c_{100}\brc{\tau} -\frac{\tau^2}{\pi^2},}
and so on. In general, (\ref{eq:Modt_N=2}) describes an important property that the functions $\hat c_{nk\brc{m}}\brc{\tau}$ are quasimodular forms of weight $2n+2k$, which was assumed in (\ref{eq:prep_N=2}) from the outset.  Moreover, these equations define the dependence of each $\hat c_{nk\brc{m}}\brc{\tau}$ on $E_2$. The only problem is that the second equation of (\ref{eq:Modt_N=2}) is quite complicated, and there is no simple way to reformulate it in terms of the recurrence relations for general $\hat c_{n}\brc{\epsilon,\tau,\hat\tau}$, as it was done in (\ref{eq:Rec_N=2}) for the modular anomaly equation (\ref{eq:mod_amomaly_6d_N=2}). Thus, we are going to use (\ref{eq:Rec_N=2}) to define the dependence of $\hat c_{n}\brc{\epsilon,\tau,\hat\tau}$ on $E_2\brc{\tau}$.
Since the recurrence relations in the $6d$ case involve an additional partial derivative with respect to the second elliptic parameter $\hat\tau$, it is useful to start with the simpler $4d$ and $5d$ cases.

\subsubsection*{Recurrence relations in $4d$}
The $4d$ limit of (\ref{eq:Rec_in_N=2}) and (\ref{eq:Rec_N=2}) can be obtained by taking $\textrm{Im}\,\hat\tau\rightarrow+\infty$ and $\epsilon\rightarrow0$:
\eqlb{eq:Rec_4d}{\begin{array}{c}
\ds 3\frac{\partial c_1}{\partial E_2} -\frac12=0,\\
\ds n\geq2:\quad 3\frac{\partial c_n}{\partial E_2}-\brc{n-1} c_{n-1}
-\frac12 \sum_{m=1}^{n-2} m\brc{n-m-1} c_{m}\, c_{n-m-1}=0,
\end{array}}
where $4d$ functions $c_n\brc{\tau}$ are related to the $6d$ functions through
\eq{c_n\brc{\tau}=\hat c_{n00}\brc{\tau}.}
Taking into account that the functions $c_n\brc{\tau}$ are quasimodular forms of weight $2n$,
we realize them as polynomials of three generators $E_2$, $E_4$ and $E_6$:
\eqlb{eq:E_gen}{\forall n\in\mathbb{N}:\quad c_n=\alpha_n\,E_2^n+\beta_{n}\,E_2^{n-2}E_4+ \gamma_{n}E_2^{n-3}E_6+\dots,}
where $\beta_{1}=0$ and $\gamma_{1}=\gamma_{2}=0$. The first coefficient $\alpha_1$ is defined by the first equation of (\ref{eq:Rec_4d}):
\eq{c_1\brc{\tau}=\frac{E_2\brc{\tau}}6.}
Then, the recurrence relations for $n\geq2$ provide us with the general expression for $\alpha_n$:
\eq{\alpha_n=\frac1{3^n\,n}\,\frac{\brc{2n-1}!!}{\brc{n+1}!}, \quad
n\geq1.}
To obtain the general expression for $\beta_n$, we use the convergency condition from section \ref{sec:mass_lim}. According to the constraints (\ref{eq:wtc_cons}), the function $\hat c_{200}\brc{\tau}$ should not contain the first power of $q$ in its expansion:
\eq{c_{2}=\alpha_2\,E_2^2+\beta_2\,E_4=\frac1{36}\,E_2^2+\beta_2\, E_4=
\frac1{36}+\beta_2+\brc{240\,\beta_2-\frac43}q+O\brc{q^2},}
which gives
\eq{\beta_2=\frac1{180}\quad \textrm{and}\quad \forall n\geq 1:\quad \beta_n=\frac25\,\frac{n-1}{3^{n}}\,\frac{\brc{2n-1}!!}{\brc{n+2}!}.}
In the same way, the general expressions for all the coefficients in (\ref{eq:E_gen}) can be obtained.
Evaluating $\gamma_n$, one gets
\eq{c_n=\frac1{3^n}\,\frac{\brc{2n-1}!!}{\brc{n+1}!}\brc{\frac1{n}\,E_2^n +
\frac25\,\frac{n-1}{n+2}\,E_2^{n-2}E_4+\frac{11}{35}\, \frac{\brc{n-1}\brc{n-2}}{\brc{n+2}\brc{n+3}}E_2^{n-3}E_6} +\dots\, .}

\subsubsection*{Recurrence relations in $5d$}
Taking $\textrm{Im}\,\hat\tau\rightarrow+\infty$, we get the $5d$ limit of (\ref{eq:Rec_in_N=2}) and (\ref{eq:Rec_N=2}):
\eqlb{eq:Rec_5d}{\begin{array}{c}
\ds 3\frac{\partial \wt c_1}{\partial E_2} -\frac12 + \epsilon^2\,\wt c_1=0,\\
\ds n\geq2:\quad
\ds 3\frac{\partial \wt c_n}{\partial E_2}-\brc{n-1}\wt c_{n-1}+ n\,\epsilon^2\,\wt c_n
-\frac12 \sum_{m=1}^{n-2} m\brc{n-m-1}\wt c_{m}\, \wt c_{n-m-1}+\\
\ds +\frac{\epsilon^2}2 \sum_{m=1}^{n-1} m\brc{n-m} \wt c_{m}\, \wt c_{n-m} =0,
\end{array}}
where
\eq{\wt c_n=\wt c_n\brc{\epsilon,\tau}=
\sum_{k=0}^{+\infty}\,\epsilon^{2k} \sum_{\at{m_1,m_2,m_3\geq 0}{m_1+2m_2+3m_3=k}}
 \hat c_{nk\brc{m}}\brc{\tau}.}
The first equation from (\ref{eq:Rec_5d}) describes an exponential dependence of $\wt c_1\brc{\epsilon,\tau}$ on $E_2$:
\eqlb{eq:c1_spdiff}{\wt c_1\brc{\epsilon,\tau}=\frac1{2\,\epsilon^2}+\alpha_1\brc{\epsilon,E_4,E_6}\, \exp\brc{-\frac{\epsilon^2}3\,E_2},}
and the unknown function $\alpha_1\brc{\epsilon,E_4,E_6}$ can be fixed by the convergency condition for the limit $\sigma\brc{\epsilon|\,\hat\tau}\rightarrow\infty$. First of all, the expansion of $\wt c_1\brc{\epsilon,\tau}$ should not contain any poles in $\epsilon$, which, along with the quasimodular properties of $\hat c_{nk0}\brc{\tau}$, provides us with the following expansion for $\alpha_1$:
\eq{\alpha_1\brc{\epsilon,E_4,E_6}=-\frac1{2\epsilon^2}\sum_{n,m\geq0}
\alpha_{1nm}\epsilon^{4n+6m}E_4^n\,E_6^m,\quad \alpha_{100}=1.}
To define other coefficients $\alpha_{1nm}$, we use the definition of the $5d$ cutoff $\wt\Lambda$:
\eqlb{eq:5d_limit}{N=2:\quad
\textrm{Im}\,\epsilon\rightarrow\infty,\quad \textrm{Im}\,\tau \rightarrow+\infty,\quad
\brc{\sin\epsilon}^2 q^{\frac12}\rightarrow\wt\Lambda^{2}.}
Then, as we established in the previous section, the prepotential should be convergent as a power series in a new parameter $\wt\nu=\sin\epsilon$, and this requirement allows one to evaluate order by order all the coefficients $\alpha_{1nm}$. However, the exact answer can be obtained, if we notice that a perfect candidate for the function convergent in the limit (\ref{eq:5d_limit}) would be the Riemann theta function. To establish the connection between the Eisenstein series and the theta functions, we consider the expansion (\ref{eq:logthetaDec2}) with $z=\epsilon$:
\eqlb{eq:dec_thetaE}{-\log\frac{\theta_{11}\brc{\pi^{-1}\epsilon|\,\tau}}
{\pi^{-1}\epsilon\,\theta_{11}'\brc{0|\,\tau}}=\sum_{k=1}^{+\infty} \frac{\zeta\brc{2k}}{k\,\pi^{2k}}\,E_{2k}\brc{\tau}\,\epsilon^{2k}.}
The theta functions at the {\it l.h.s.} of (\ref{eq:dec_thetaE}) are automatically convergent in the limit (\ref{eq:5d_limit}), if we appropriately rescale the first few terms in their expansion in powers of $q$.
At the same time, the {\it r.h.s.} of (\ref{eq:dec_thetaE}) tells us how to apply these theta functions to (\ref{eq:c1_spdiff}). The first term in the expansion (\ref{eq:dec_thetaE}) is $\epsilon^2\, E_2/6$, which gives the exact answer for the function $\wt c_1\brc{\epsilon,\tau}$:
\eqlb{eq:c1_exact}{\boxed{\wt c_1\brc{\epsilon,\tau}=\frac1{2\,\epsilon^2}-\frac{\pi^2}{2\,\epsilon^4}\, \frac{\theta_{11}\brc{\pi^{-1}\epsilon|\,\tau}^2}
{\theta_{11}'\brc{0|\,\tau}^2}.}}
The combination $\epsilon^4\,\wt c_1\brc{\epsilon,\tau}$ is just the theta function up to some $q$-independent shift, and the first few terms in the $q$-expansion are
\eq{\epsilon^4\,\wt c_1\brc{\epsilon,\tau}=\frac{\epsilon^2}{2}-\frac12\brc{\sin\epsilon}^2- 4\,q\brc{\sin\epsilon}^4- 4\,q^2\brc{3+2\brc{\sin\epsilon}^2}\brc{\sin\epsilon}^4+ O\brc{q^3}.}
An additional check of (\ref{eq:c1_exact}) is provided by the perturbative
limit $\textrm{Im}\,\tau \rightarrow+\infty$:
\eq{\epsilon^4\,\wt c_1\brc{\epsilon,+\infty}=\frac{\epsilon^2}2-\frac12\brc{\sin\epsilon}^2,}
which matches the exact answer for the perturbative part of the prepotential (\ref{eq:pert_N=2}).

The equation for the second function $\wt c_2\brc{\epsilon,\tau}$ is
\eq{3\frac{\partial \wt c_2}{\partial E_2}+ 2\,\epsilon^2\,\wt c_2-\wt c_{1}
+\frac{\epsilon^2}2\,\wt c_{1}^{\,2}=0.}
Substituting $\wt c_1$ in the form (\ref{eq:c1_spdiff}), we get the following solution:
\eq{\wt c_2\brc{\epsilon,\tau}=\frac{3}{16\,\epsilon^4}+\frac{\alpha_1}{2\,\epsilon^2} \exp\brc{-\frac{\epsilon^2}3\,E_2} + \brc{ \alpha_2\brc{\epsilon,E_4,E_6} -\frac{\epsilon^2}{6}\,E_2\,\alpha_1^2} \exp\brc{-\frac{2\,\epsilon^2}3\,E_2},}
which can be rewritten with the help of (\ref{eq:c1_exact}) as follows:
\eqlb{eq:c2_spdiff}{\wt c_2\brc{\epsilon,\tau} =\frac{3}{16\,\epsilon^4}-\frac{\pi^2}{4\,\epsilon^6}
\frac{\theta_{11}\brc{\pi^{-1}\epsilon|\,\tau}^2} {\theta_{11}'\brc{0|\,\tau}^2} +\frac{\pi^4}{8\,\epsilon^6} \frac{\theta_{11}\brc{\pi^{-1}\epsilon|\,\tau}^4} {\theta_{11}'\brc{0|\,\tau}^4}\brc{\frac1{2\,\epsilon^2}-\frac1{3}E_2+ \beta_1\brc{\epsilon,E_4,E_6}}.}
Due to the quasimodular properties, the unknown function $\beta_1$ has the following expansion:
\eq{\beta_1\brc{\epsilon,E_4,E_6}=\frac1{\epsilon^2}\sum_{\at{n,m\geq0}{n+m>0}}
\beta_{1nm}\epsilon^{4n+6m}E_4^n\,E_6^m.}
This function is nontrivial, because in (\ref{eq:c2_spdiff}) there is the term
\eq{\pi^4\frac{\theta_{11}\brc{\pi^{-1}\epsilon|\,\tau}^4}
{\theta_{11}'\brc{0|\,\tau}^4} =\brc{\sin\epsilon}^4+16\,q\brc{\sin\epsilon}^6+ 48\,q^2\brc{1+2\brc{\sin\epsilon}^2}\brc{\sin\epsilon}^6+ O\brc{q^3},}
which leads at generic $\beta_1$ to the divergence of $\epsilon^6\,\wt c_2$ when the imaginary part of $\epsilon$ goes to infinity. In fact, there are two divergent terms: the perturbative one $\brc{\sin\epsilon}^4$ and non-perturbative $q\brc{\sin\epsilon}^6\sim \brc{\sin\epsilon}^2$. The former one is separately dealt with in (\ref{eq:pert_N=2}), and the perturbative limit of $\wt c_2$ is given by
\eqlb{eq:pert_c2}{\epsilon^6\,\wt c_2\brc{\epsilon,+\infty}=\frac1{16} \brc{3\,\epsilon^2-3\brc{\sin\epsilon}^2-\brc{\sin\epsilon}^4}}
This fixes only the initial condition for $\beta_1$:
\eqlb{eq:pert_b1}{\frac1{2\,\epsilon^2}+\beta_1\brc{\epsilon,1,1}=-\frac16+\frac1{2\brc{\sin\epsilon}^2}.}
This gives a hint that $\beta_1$ can be expressed in terms of the theta functions, which should be checked by the requirement of cancelling the non-perturbative $q\brc{\sin\epsilon}^6\sim \brc{\sin\epsilon}^2$ divergency. Indeed, consider the second partial derivative of (\ref{eq:dec_thetaE}) with respect to $\epsilon$:
\eq{-\partial_{\epsilon}^2 \log\theta_{11}\brc{\pi^{-1}\epsilon|\,\tau} -
\frac1{\epsilon^2} = \frac13 E_2+2 \sum_{k=2}^{+\infty} \frac{\zeta\brc{2k}}{\pi^{2k}}
\brc{2k-1} E_{2k}\brc{\tau}\,\epsilon^{2k-2},}
or, equivalently,
\eq{\frac1{2\,\epsilon^2}+\sum_{k=2}^{+\infty} \frac{\zeta\brc{2k}}{\pi^{2k}}
\brc{2k-1} E_{2k}\brc{\tau}\,\epsilon^{2k-2} =- \frac16 E_2 -\frac12 \partial_{\epsilon}^2 \log\theta_{11}\brc{\pi^{-1}\epsilon|\,\tau}.}
The {\it r.h.s.} of this latter equation perfectly fits the perturbative limit (\ref{eq:pert_b1})
and the {\it l.h.s.} provides the series expansion for $\beta_1$. Making sure that the first few terms in the expansion of $\epsilon^6\,\wt c_2$ are convergent, we get the exact expression
\eqlb{eq:c2_exact}{\boxed{\wt c_2\brc{\epsilon,\tau} =\frac{3}{16\,\epsilon^4}-\frac{\pi^2}{4\,\epsilon^6}
\frac{\theta_{11}\brc{\pi^{-1}\epsilon|\,\tau}^2} {\theta_{11}'\brc{0|\,\tau}^2} -\frac{\pi^4}{16\,\epsilon^6} \frac{\theta_{11}\brc{\pi^{-1}\epsilon|\,\tau}^4} {\theta_{11}'\brc{0|\,\tau}^4}\brc{E_2\brc{\tau}- \frac1{\sigma\brc{\epsilon|\,\tau}^2}}.}}
In the same way, the recurrence relations (\ref{eq:Rec_5d}) allow one to evaluate the functions $\wt c_n$ at any given $n$.

\subsubsection*{Recurrence relations in $6d$}
Finally, we come to the most general $6d$ case and to the recurrence relations (\ref{eq:Rec_in_N=2}) and (\ref{eq:Rec_N=2}). The first equation (\ref{eq:Rec_in_N=2}) includes both functions $\hat c_{1}\brc{\epsilon,\tau,\hat\tau}$ and  $\hat c_{2}\brc{\epsilon,\tau,\hat\tau}$. However, the second function appears in terms of higher orders in $\epsilon$. The same is true for other equations (\ref{eq:Rec_N=2}), where the function $\hat c_{n+1}\brc{\epsilon,\tau,\hat\tau}$ appears in  $\epsilon^4$- and $\epsilon^6$-terms. This allows one to calculate the coefficients $\hat c_{nk\brc{m}}\brc{\tau}$ in the series expansions (\ref{eq:cn_Dec}) order by order.
Using the convergency condition for the limit $\sigma\brc{\epsilon|\,\hat\tau}\rightarrow\infty$, we compute some first coefficients in the $\epsilon$-expansion of the functions $\hat c_{n}\brc{\epsilon,\tau,\hat\tau}$:
\eq{\def\arraystretch{2.2}
\begin{array}{c}
\ds\hat c_{1} \brc{\epsilon,\tau,\hat\tau}= \frac{E_2}6- \epsilon^2 \frac{\hat E_2}{180} \brc{5\,E_2^2-E_4} +\epsilon^4 \frac{\hat E_2^2}{13\,608} \brc{ 70\,E_2^3 - 21\,E_2\, E_4 + 5\,E_6}-\\
\ds -\epsilon^4 \frac{\hat E_4}{68\,040} \brc{140\,E_2^3 + 21\,E_2\,E_4 + E_6}+O\brc{\epsilon^6},
\end{array}}
\eq{\def\arraystretch{2.2}
\begin{array}{c}
\ds \hat c_{2} \brc{\epsilon,\tau,\hat\tau}=\frac1{180}\brc{5\,E_2^2+E_4}- \epsilon^2 \frac{\hat E_2}{22\,680} \brc{245\,E_2^3 + 42\,E_2\,E_4 - 17\,E_6} +\epsilon^4 \frac{\hat E_2^2}{136\,080} \left( 455\,E_2^4 + 91\,E_2^2\, E_4 -\right.\\
\ds \left.- 46\,E_2\,E_6+4\,E_4^2\right) - \epsilon^4 \frac{\hat E_4}{1\,360\,800} \brc{1225\,E_2^4 + 700\,E_2^2\, E_4 + 60\,E_2\,E_6+31\,E_4^2} + O\brc{\epsilon^6},
\end{array}}
\eq{\def\arraystretch{2.2}
\begin{array}{c}
\ds \hat c_{3} \brc{\epsilon,\tau,\hat\tau}=\frac1{22\,680}\brc{175\,E_2^3 + 84\,E_2\,E_4 + 11\,E_6}- \epsilon^2 \frac{\hat E_2}{136\,080} \brc{665\,E_2^4 + 357\,E_2^2\, E_4 - 2\,E_2\,E_6-12\,E_4^2} +\\
\ds+ \frac{\epsilon^4} {8\,981\,280} \left(\hat E_2^2 \brc{19\,250\,E_2^5 + 12\,089\,E_2^3\, E_4 - 759\,E_2^2\,E_6-319\,E_2\,E_4^2-21\,E_4\,E_6} -\right.\\
\ds \left. - \hat E_4\brc{3\,927\,E_2^5 + 3\,773\,E_2^3\, E_4 + 737\,E_2^2\,E_6+ 572\,E_2\,E_4^2+ 63\,E_4\,E_6}\right) + O\brc{\epsilon^6},
\end{array}}
where the notation is usual: $E_{2k}\equiv E_{2k}\brc{\tau}$ and $\hat E_{2k}\equiv E_{2k}\brc{\hat\tau}$. It can be easily checked that the above expressions are in complete agreement with the recurrence relations (\ref{eq:mod_anomaly2_Rec1}) coming from the second modular anomaly equation (\ref{eq:mod_amomaly_2}). In other words, each function $\hat c_{n}\brc{\epsilon,\tau,\hat\tau}$ with $n\geq2$ can be obtained from $\hat c_{1}\brc{\epsilon,\tau,\hat\tau}$ with the help of relation (\ref{eq:mod_anomaly2_Rec2}):
\eq{n\geq2:\quad
\hat c_n=\frac{\brc{-1}^{n-1}}{n!}
\brc{\frac3{\epsilon^2}\frac{\partial}{\partial \hat E_2} +\frac1{\pi\imath} \frac{\partial}{\partial \tau}}^{n-1}\hat c_1.}
This claim is supported by the first $22$ orders in the $\epsilon$-expansion of $\hat c_{1}\brc{\epsilon,\tau,\hat\tau}$. Thus, provided the second modular anomaly equation is correct, the computation of the $N=2$ double-elliptic prepotential reduces to finding just the first function $\hat c_{1}$. To this end, the two modular anomaly equations can be combined. Using (\ref{eq:mod_anomaly2_Rec2}) with $n=2$ and equation (\ref{eq:Rec_in_N=2}), one gets the first-order partial differential equation for the function $\hat c_{1}$, and the boundary condition for this equation is given by the limit $\sigma\brc{\epsilon|\,\hat\tau}\rightarrow\infty$.

To conclude the $N=2$ part of the paper, let us discuss some important properties of the function $\hat c_{1}\brc{\epsilon,\tau,\hat\tau}$, in particular, look at the series expansions in powers of the other two parameters $q=\exp\brc{2\pi\imath\,\tau}$ and $\hat q=\exp\brc{2\pi\imath\,\hat\tau}$. Relying on the computed orders in the $\epsilon$-expansion of $\hat c_{1}$, we establish the first few orders in the $q$-expansion:
\eqlb{eq:c1_q1Exp}{\def\arraystretch{2.2}
\begin{array}{c}
\ds \hat c_{1}\brc{\epsilon,\tau,\hat\tau}=\hat c_{1}^{\textrm{pert}}\brc{\epsilon,\hat\tau}- q \frac{4\,\pi^{4}}{\epsilon^4} \frac{\theta_{11}\brc{\pi^{-1}\epsilon|\,\hat\tau}^{4}} {\theta_{11}'\brc{0|\,\hat\tau}^{4}}-
q^2 \frac{2\,\pi^{8}}{\epsilon^4} \frac{\theta_{11}\brc{\pi^{-1}\epsilon|\,\hat\tau}^{8}} {\theta_{11}'\brc{0|\,\hat\tau}^{8}} \left(\frac{6}{\sigma\brc{\epsilon|\,\hat\tau}^4} +\right.\\
\ds\left. + \frac{4\,\hat E_2}{\sigma\brc{\epsilon|\,\hat\tau}^2}-\frac{\hat E_2^2-\hat E_4}3\right)+ O\brc{q^3},
\end{array}}
where
\eq{c_n^{\textrm{pert}}\brc{\epsilon,\hat\tau}=\lim_{\textrm{Im}\,\tau\rightarrow+\infty} \hat c_{n}\brc{\epsilon,\tau,\hat\tau}}
The expansion (\ref{eq:c1_q1Exp}) is consistent with the symmetry (\ref{eq:esym6d}) and justifies the choice of the parameters $\wt q$ and $\nu$ in Section \ref{sec:mass_lim}. The function $\hat c_{1}^{\textrm{pert}} \brc{\epsilon,\hat\tau}$ contributes to the exact expression for the perturbative part of the period matrix (\ref{eq:pert_N=2}). With the help of (\ref{eq:pert_N=2}), we derive the following equation:
\eq{\brc{\hat E_2^3-3\,\hat E_2\,\hat E_4+2\hat E_6}\frac{\partial\,\hat c_{1}^{\textrm{pert}}}{\partial \hat E_2}+\frac32\brc{\hat E_2^2-
\hat E_4}\hat c_{1}^{\textrm{pert}} + \frac{9}{\epsilon^4} \log\frac{\pi\,\theta_{11}\brc{\pi^{-1}\epsilon|\,\hat\tau}}
{\sigma\brc{\epsilon|\,\hat\tau}\theta_{11}'\brc{0|\,\hat\tau}}=0,}
which can be used to calculate the function $\hat c_{1}^{\textrm{pert}} \brc{\epsilon,\hat\tau}$ up to any given order in $\epsilon$.

The first few orders in the $\hat q$-expansion of the function $\hat c_{1}$ are
\eq{\def\arraystretch{2.2}
\begin{array}{c}
\ds \hat c_{1}\brc{\epsilon,\tau,\hat\tau}= \frac1{2\,\epsilon^2}-
\frac{\pi^2}{2\,\epsilon^4}\, \frac{\theta_{11}\brc{\pi^{-1}\epsilon|\,\tau}^2}
{\theta_{11}'\brc{0|\,\tau}^2}+\frac{2\,\hat q}{\epsilon^4} \brc{-\epsilon^2+2\pi^2 \frac{\theta_{11}\brc{\pi^{-1}\epsilon|\,\tau}^2} {\theta_{11}'\brc{0|\,\tau}^2}} +\\
\ds +\frac{2\,\hat q}{\epsilon^4}\brc{\pi^4 \frac{\theta_{11}\brc{\pi^{-1}\epsilon|\,\tau}^4} {\theta_{11}'\brc{0|\,\tau}^4}\brc{E_2- \frac1{\sigma\brc{\epsilon|\,\tau}^2}}+ \frac{2\,\pi^6}9 \frac{\theta_{11}\brc{\pi^{-1}\epsilon|\,\tau}^6} {\theta_{11}'\brc{0|\,\tau}^6}\brc{2\,E_2^2- \frac{3\,E_2}{\sigma\brc{\epsilon|\,\tau}^2}}} + O\brc{\hat q^2}.
\end{array}}
As expected, the zeroth order is the $5d$ function $\wt c_1\brc{\epsilon,\tau}$ defined in (\ref{eq:c1_exact}). The structures of the both $q$- and $\hat q$-expansions are similar: the coefficients are specific theta functions of $\epsilon$ with moduli $\hat\tau$ and $\tau$ correspondingly. Thus, the exact expressions at any finite order in these expansions can be computed.


\section{Modular anomaly at $N=3$}
\label{sec:N=3}
In this section, we use the modular anomaly equations in order to compute a few first orders in the $\epsilon$-expansion of the functions $\widehat{C}_{i_1,\dots,i_n}\brc{\epsilon,\tau,\hat\tau}$ in the case of $N=3$. Since $\widehat{C}_{i_1,\dots,i_n}$ are fully symmetric under the permutation of indices $i_1,\dots,i_n$, we introduce the variables
\eq{\hat s_{i_1,i_2,i_3}\brc{\beta\,\textbf{a},\hat\tau}= \frac12\sum_{\at{\vec\alpha_1,\vec\alpha_2,\vec\alpha_3 \in\Delta_+}{\vec\alpha_1\neq\vec\alpha_2\neq\vec\alpha_3}} \frac1{\sigma\brc{\beta\,\vec\alpha_1\cdot\vec a|\,\hat\tau}^{2\,i_1}} \frac1{\sigma\brc{\beta\,\vec\alpha_2\cdot\vec a|\,\hat\tau}^{2\,i_2}} \frac1{\sigma\brc{\beta\,\vec\alpha_3\cdot\vec a|\,\hat\tau}^{2\,i_3}}}
with $\Delta_+=\bfi{\vec e_1-\vec e_2,\,\vec e_1-\vec e_3,\,\vec e_2-\vec e_3}$ and $\vec e_i \cdot \vec a=a_i$. Then, the prepotential (\ref{eq:prepDellN}) can be written at $N=3$ as
\eqlb{eq:prepDellN=3}{\begin{array}{c}\ds
\mathcal{F}^{\textrm{Dell}}=\frac{\tau}2\sum_{i=1}^{3}a_i^2+
\frac{\epsilon^2}{2\pi\imath\,\beta^2}
\sum_{i<j}\log\,\theta_{11}\brc{\pi^{-1}\beta\,a_{ij}|\,\hat\tau}-\\
\ds
-\frac{\epsilon^2}{\pi\imath\,\beta^2}\sum_{n\in\mathbb{N}}\sum_{i=0}^{1}\sum_{j=i}^{n}\, \epsilon^{2\brc{n+n i+ j}}\, \widehat{C}_{n,n i,j}\brc{\epsilon,\tau,\hat\tau}\,
\hat s_{n,n i,j}\brc{\beta\,\textbf{a},\hat\tau},
\end{array}}
where again the summation over the index $i$ was restricted, since otherwise not all the coefficients $\widehat{C}_{i_1,i_2,i_3}\brc{\epsilon,\tau,\hat\tau}$ would be independent because of the relations between $\sigma\brc{\beta\,\vec\alpha_k\cdot\vec a\,|\,\hat\tau}$ and, hence, between $\hat s_{i_1,i_2,i_3}\brc{\beta\,\textbf{a},\hat\tau}$.

With the help of the first modular anomaly equation (\ref{eq:mod_amomaly_6d_N}) and of the convergency condition (\ref{eq:wtc_cons}), we compute the coefficients in the prepotential (\ref{eq:prepDellN=3}) up to the order of $\epsilon^{14}$. This corresponds to computing quasimodular forms $\widehat{C}_{i_1,i_2,i_3,k,\brc{m}}\brc{\tau}$ in the expansions (\ref{eq:C_Dec_N}) up to the weight $12$. For a few first functions $\widehat{C}_{i_1,i_2,i_3}\brc{\epsilon,\tau,\hat\tau}$, the expansions are
\eq{\def\arraystretch{2.2}
\begin{array}{c}
\ds \widehat{C}_{100}\brc{\epsilon,\tau,\hat\tau}= \frac{E_2}6- \epsilon^2 \frac{\hat E_2}{180} \brc{5\,E_2^2-E_4} +\epsilon^4 \frac{\hat E_2^2}{7\,560} \brc{35\,E_2^3 - 7\,E_2\, E_4 + 2\,E_6}- \\ \ds
-\epsilon^4 \frac{\hat E_4}{22\,680} \brc{35\,E_2^3 + 21\,E_2\,E_4 -2 E_6}+O\brc{\epsilon^6},
\end{array}}
\eq{\def\arraystretch{2.2}
\begin{array}{c}
\ds \widehat{C}_{101}\brc{\epsilon,\tau,\hat\tau}= -\frac1{144}\brc{E_2^2-E_4}+ \epsilon^2 \frac{\hat E_2}{6\,480} \brc{25\,E_2^3 -33\,E_2\,E_4 +8\,E_6} -\epsilon^4 \frac{\hat E_2^2}{2\,177\,280} \left( 3535\,E_2^4 -\right.\\
\ds \left.- 5082\,E_2^2\, E_4 + 1592\,E_2\,E_6- 45\,E_4^2\right) + \epsilon^4 \frac{\hat E_4}{725\,760} \brc{385\,E_2^4 - 294\,E_2^2\, E_4 - 88\,E_2\,E_6 - 3\,E_4^2} + O\brc{\epsilon^6},
\end{array}}
\eq{\def\arraystretch{2.2}
\begin{array}{c}
\ds \widehat{C}_{200}\brc{\epsilon,\tau,\hat\tau}= \frac1{180}\brc{5\,E_2^2+E_4}- \epsilon^2 \frac{\hat E_2}{22\,680} \brc{245\,E_2^3 + 42\,E_2\,E_4 - 17\,E_6} + \epsilon^4 \frac{\hat E_2^2}{2\,177\,280} \left(6965\,E_2^4 + \right.\\
\ds \left. + 1722\,E_2^2\, E_4 - 632\,E_2\,E_6+ 9\,E_4^2\right) - \epsilon^4 \frac{\hat E_4}{10\,886\,400} \brc{8225\,E_2^4 + 6930\,E_2^2\, E_4 + 1000\,E_2\,E_6 - 27\,E_4^2} + O\brc{\epsilon^6},
\end{array}}
and so on. Then, the first few orders in the $q$-expansion of the function $\widehat{C}_{100}$ can be established:
\eqlb{eq:C100_q1Exp}{\def\arraystretch{2.2}
\begin{array}{c}
\ds \widehat{C}_{100}\brc{\epsilon,\tau,\hat\tau}=  \widehat{C}^{\textrm{pert}}_{100}\brc{\epsilon,\hat\tau}- q \frac{\pi^{6}}{\epsilon^4} \frac{\theta_{11}\brc{\pi^{-1}\epsilon|\,\hat\tau}^{6}} {\theta_{11}'\brc{0|\,\hat\tau}^{6}} \frac{1}{\sigma\brc{\epsilon|\,\hat\tau}^2} - q^2 \frac{\pi^{12}}{\epsilon^4} \frac{\theta_{11}\brc{\pi^{-1}\epsilon|\,\hat\tau}^{12}} {\theta_{11}'\brc{0|\,\hat\tau}^{12}} \left[\frac{3}{\sigma\brc{\epsilon|\,\hat\tau}^8} +\right.\\
\ds\left. + \frac{2\,\hat E_2}{\sigma\brc{\epsilon|\,\hat\tau}^6}+\frac12\brc{\hat E_2^2-\hat E_4} \frac1{\sigma\brc{\epsilon|\,\hat\tau}^4} -\frac{10}{27}\brc{\hat E_2^3-3\,\hat E_2\,\hat E_4+2\, \hat E_6} \frac1{\sigma\brc{\epsilon|\,\hat\tau}^2} +\right. \\
\ds\left. + \frac{1}{27}\brc{7\,\hat E_2^4-18\,\hat E_2^2\,\hat E_4+8\,\hat E_2\,\hat E_6+3\,\hat E_4^2\,} \right]  + O\brc{q^3}.
\end{array}}

We checked that the second modular anomaly equation (\ref{eq:mod_amomaly_N2}) works in all computed orders. For example, (\ref{eq:mod_amomaly_N2}) provides us with the relation
\eq{\frac1{\pi\imath}\frac{\partial \widehat{C}_{100}}{\partial \tau}+ \frac2{\epsilon^2}\frac{\partial \widehat{C}_{100}}{\partial\hat E_2} +\frac43\brc{\widehat{C}_{101}+\widehat{C}_{200}}=O\brc{\epsilon^6},}
where the {\it r.h.s.} is non-zero due to the presence of other functions $\widehat{C}_{i_1,i_2,i_3}\brc{\epsilon,\tau,\hat\tau}$ at higher orders in $\epsilon$.

The results of this section are in complete agreement with the calculations from \cite{AMM'16}, where the involutivity conditions for the double-elliptic Hamiltonians were used to compute the corresponding Seiberg-Witten prepotential. To compare the prepotential in the form  (\ref{eq:prepDellN=3}) and the prepotential from \cite{AMM'16}, one should use the following transition formula:
\eq{\frac1{\sigma\brc{z|\,\hat\tau}^2}= \frac1{\sn\brc{z|\,\hat\tau}^2}-\vartheta\brc{\hat\tau},}
where the function $\sn\brc{z|\,\hat\tau}$ is just the rescaled Jacobi elliptic function:
\eq{\sn\brc{z|\,\hat\tau}=
\pi \frac{\theta_{01}\brc{0|\,\hat{\tau}}}{\theta'_{11}\brc{0|\,\hat{\tau}}}\,
\frac{\theta_{11}\brc{\pi^{-1} z|\,\hat{\tau}}}{\theta_{01}\brc{\pi^{-1} z|\,\hat{\tau}}}}
and the function $\vartheta\brc{\hat\tau}$ is
\eq{\vartheta\brc{\hat\tau}=\frac{4\imath}{\pi}\,
\partial_{\hat\tau}\log\theta_{01}\brc{0|\,\hat{\tau}}.}
Moreover, the $\epsilon$-expansion of the prepotential in \cite{AMM'16} is written in terms of the theta constants $\vartheta\brc{\hat\tau}$, $\theta_{00}\brc{0|\,\hat{\tau}}^4$, and $\theta_{10}\brc{0|\,\hat{\tau}}^4$. To rewrite this expansion in terms of the Eisenstein series $E_2\brc{\hat\tau}$, $E_4\brc{\hat\tau}$, and $\hat E_6\brc{\hat\tau}$, we use the theta-constant identities of the form
\eq{E_2\brc{\hat\tau}=\theta_{00}\brc{0|\,\hat{\tau}}^4+\theta_{10}\brc{0|\,\hat{\tau}}^4-3\,\vartheta\brc{\hat\tau},}
\eq{E_4\brc{\hat\tau}=\theta_{00}\brc{0|\,\hat{\tau}}^8+\theta_{10}\brc{0|\,\hat{\tau}}^8 - \theta_{00}\brc{0|\,\hat{\tau}}^4\theta_{10}\brc{0|\,\hat{\tau}}^4,}
\eq{E_6\brc{\hat\tau}=\theta_{00}\brc{0|\,\hat{\tau}}^{12}+\theta_{10}\brc{0|\,\hat{\tau}}^{12}-\frac32\, \theta_{00}\brc{0|\,\hat{\tau}}^4\theta_{10}\brc{0|\,\hat{\tau}}^4 \brc{\theta_{00}\brc{0|\,\hat{\tau}}^4+\theta_{10}\brc{0|\,\hat{\tau}}^4}.}

\section{Conclusion}
We discussed a number of questions concerning the low energy effective action of the $6d$ SYM theory with two compactified Kaluza-Klein dimensions and the adjoint matter hypermultiplet. The main focus of our study was on the properties of the theory under the modular transformations of the two elliptic parameters $\tau$ and $\hat\tau$. As a result, two modular anomaly equations were derived, and the corresponding duality group of the theory was described by four generators (\ref{eq:sym_gen}). We demonstrated that the first modular anomaly equation (\ref{eq:mod_amomaly_6d_N}) provides a new method to compute the double-elliptic Seiberg-Witten prepotential, if proper boundary conditions are imposed. For small enough values of $N$, such boundary conditions are given by the limit $\sigma\brc{\epsilon|\,\hat\tau}\rightarrow\infty$. This method of calculating the Seiberg-Witten prepotentials is rather simple and could help to achieve further advance in study of the double-elliptic integrable systems.

Of course, there are still many problems to investigate. In particular, the curve $\Gamma^{\textrm{Dell}}$ lacks any clear geometrical description that would manifest all the basic properties discussed in this paper. There were different attempts in this direction, including an interpretation of $\Gamma^{\textrm{Dell}}$ as a complex Riemann surface of genus $N+1$ in \cite{Braden2003} and the theta-constant representation for the Seiberg-Witten curves in \cite{ABMMZ}. In section \ref{sec:First_MA}, we also gained some clues on how the curve should be described. Nonetheless, all these pieces of data are quite unrelated, and a separate effort needs to be done to put it all together. An interpretation of the obtained results at the level of Nekrasov partition functions could also be an interesting research direction. As it was mentioned in the Introduction, we are almost certain that there is an uplift of the $6d$ modular anomaly equations to the level of Nekrasov functions. What would happen to other properties such as the symmetry (\ref{eq:esym6d}) is not that clear.

\section*{Acknowledgements}
We are grateful to Y. Zenkevich for helpful discussions. The work was partly supported by the grant of the Foundation for the Advancement of Theoretical Physics ``BASIS" (A.Mor.), by grant 16-32-00920-mol-a (G.A.), by RFBR grants 15-02-04175 (G.A.), 16-01-00291 (A.Mir.) and 16-02-01021 (A.Mor.), by joint grants 17-51-50051-YaF (A.Mir. and A.Mor.), 15-51-52031-NSC-a, 16-51-53034-GFEN, 16-51-45029-IND-a.

\appendix
\section{$\sigma$ function}
\label{sec:sigma}
The $\sigma$ function is directly connected with the Weierstrass $\wp$ function:
\eqlb{eq:appsigmaW}{\frac1{\sigma\brc{z|\,\hat{\tau}}^2}\equiv-\partial_{z}^2\log\theta_{11}\brc{\pi^{-1}z|\,\hat{\tau}}= \frac{1}{3}E_2\brc{\hat\tau}+\wp\brc{z|\,\hat{\tau}}.}
Some standard differential equations for $\sigma\brc{z|\,\hat{\tau}}$ are
\eq{\sigma'\brc{z|\,\hat\tau}^2=1-\hat E_2\,\sigma^2+\frac13\brc{\hat E_2^2-\hat E_4}\sigma^4 - \frac1{27}\brc{\hat E_2^3-3\hat E_2\hat E_4+2\hat E_6}\sigma^6,}
\eq{\sigma''\brc{z|\,\hat\tau}=-\hat E_2\,\sigma+\frac23\brc{\hat E_2^2-\hat E_4}\sigma^3 - \frac1{9}\brc{\hat E_2^3-3\hat E_2\hat E_4+2\hat E_6}\sigma^5,}
\eq{\frac2{\pi\imath}\partial_{\hat\tau}\sigma\brc{z|\,\hat\tau} = \frac1{\sigma} - \hat E_2\, \sigma+\frac16\brc{\hat E_2^2-\hat E_4}\sigma^3 - \sigma'\brc{z|\,\hat\tau}
\partial_z\log\theta_{11}\brc{\pi^{-1}z|\,\hat\tau},}
where we use the notation
\eq{\sigma'\brc{z|\,\hat\tau}=\partial_z\sigma\brc{z|\,\hat{\tau}}.}

\bibliographystyle{unsrt}
\bibliography{references}

\end{document}